\documentclass[usenatbib]{mn2e}
\usepackage{amsmath}
\usepackage{natbib}
\usepackage{epsfig}
\usepackage{txfonts}

\newcommand{\vbetac}{{{\boldsymbol\beta}_{\rm c}}}
\newcommand{\vbetach}{{\hat{{\boldsymbol\beta}}_{\rm c}}}
\newcommand{\betac}{\beta_{\rm c}}

\newcommand{\betacsq}{{\beta^2_{\rm c}}}

\newcommand{\muc}{\mu_{\rm c}}

\newcommand{\vbetao}{{{\boldsymbol\beta}_{\rm o}}}
\newcommand{\vbetaoh}{{\hat{{\boldsymbol\beta}}_{\rm o}}}
\newcommand{\betao}{\beta_{\rm o}}
\newcommand{\muo}{\mu_{\rm o}}

\newcommand{\vecJ}[1]{{\bf #1}}

\newcommand{\vgh}{{\hat{\boldsymbol\gamma}}}
\newcommand{\vghp}{{\hat{\boldsymbol\gamma}'}}

\newcommand{\vbh}{{\boldsymbol{\hat{\beta}}}}

\newcommand{\vp}{{\vecJ{p}}}
\newcommand{\vpp}{{\vecJ{p'}}}

\newcommand{\Deltaxe}{\Delta \xe}

\newcommand{\xg}{x}

\newcommand{\xgc}{x_{\rm c}}

\newcommand{\xe}{x_{\rm e}}

\newcommand{\id}{{\,\rm d}}

\newcommand{\beq}{\begin{equation}}   %

\newcommand{\eeq}{\end{equation}}   %

\newcommand{\beqa}{\begin{eqnarray}}   %

\newcommand{\eeqa}{\end{eqnarray}}   %

\newcommand{\beal}{\begin{align}}

\newcommand{\enal}{\end{align}}

\newcommand{\bspl}{\begin{split}}

\newcommand{\espl}{\end{split}}

\newcommand{\bsub}{\begin{subequations}}

\newcommand{\esub}{\end{subequations}}

\newcommand{\bmulti}{\begin{multline}}   %

\newcommand{\beqm}{\begin{mathletters}}   %

\newcommand{\eeqm}{\end{mathletters}}   %

\newcommand{\Abst}[1]{\,#1}

\newcommand{\me}{m_{\rm e}}

\newcommand{\Ne}{N_{\rm e}}

\newcommand{\Te}{T_{\rm e}}

\newcommand{\Tg}{T_{\gamma}}

\newcommand{\The}{\theta_{\rm e}}

\newcommand{\sigT}{\sigma_{\rm T}}

\newcommand{\nPl}{n_{\rm Pl}}

\newcommand{\vek} [1]{\mbox{\boldmath${#1}$\unboldmath}}

\newcommand{\pd}{\partial}

\newcommand{\pAb}[2]{\frac{\displaystyle\pd #1}{\displaystyle\pd #2}}

\newcommand{\pot}[2]{#1 \times 10^{#2}}

\usepackage{color}

\usepackage{hyperref}
\usepackage{grffile}
\usepackage{graphics}
\usepackage{booktabs}

\title[Computation of SZ signal]
{A fast and accurate method for computing the Sunyaev-Zeldovich signal of hot galaxy clusters}

\author[Chluba et al.]{Jens Chluba$^{1}$\thanks{E-mail:
  jchluba@cita.utoronto.ca}, 
  Daisuke Nagai$^{2, 3, 4}$, Sergey Sazonov$^{5, 6}$, Kaylea Nelson$^{3}$
  \\
  \\
$^{1}$ Canadian Institute for Theoretical Astrophysics, 60 St. George Street,
Toronto, ON M5S 3H8, Canada
\\
$^{2}$ Department of Physics, Yale University, New Haven, CT 06520, U.S.A.
\\
$^{3}$ Department of Astronomy, Yale University, New Haven, CT 06520, U.S.A.
\\
$^{4}$ Yale Center for Astronomy \& Astrophysics, Yale University, New Haven, CT 06520, U.S.A.
\\
$^{5}$ Space Research Institute, Russian Academy of Sciences, Profsoyuznaya 84/32, 117997 Moscow, Russia 
\\
$^{6}$ Max-Planck-Institut f{\"u}r Astrophysik, Karl-Schwarzschild-Str. 1, 85740 Garching bei M{\"u}nchen, Germany
}

\begin{document}

\date{Received 2012 May 24; Accepted 2012 July 18}

\maketitle

\begin{abstract}
New generation ground and space-based CMB experiments have ushered in discoveries of massive galaxy clusters via the Sunyaev-Zeldovich (SZ) effect, providing a new window for studying cluster astrophysics and cosmology. 
Many of the newly discovered, SZ-selected clusters contain hot intracluster plasma  ($k\Te\gtrsim 10\,{\rm keV}$) and exhibit disturbed morphology, indicative of frequent mergers with large peculiar velocity ($v \gtrsim 1000\, {\rm km s}^{-1}$). 
It is well-known that for the interpretation of the SZ signal from hot, moving galaxy clusters, relativistic corrections must be taken into account,
and in this work, we present a fast and accurate method for computing these effects. 
Our approach is based on an alternative derivation of the Boltzmann collision term which provides new physical insight into the sources of different kinematic corrections in the scattering problem. 
This allows us to obtain a clean separation of kinematic and scattering terms which differs from previous works.
We also briefly mention additional complications connected with kinematic effects  that should be considered when interpreting future SZ data for individual clusters.
One of the main outcomes of this work is {\sc SZpack}, a numerical library which allows very fast and precise ($\lesssim 0.001\%$ at frequencies $h\nu \lesssim 20 k\Tg$) computation of the SZ signals up to high electron temperature ($k\Te \simeq 25\,{\rm keV}$) and large peculiar velocity ($v/c \simeq 0.01$).
The accuracy is well beyond the current and future precision of SZ observations and practically eliminates uncertainties related to more expensive numerical evaluation of the Boltzmann collision term.
Our new approach should therefore be useful for analyzing future high-resolution, multi-frequency SZ observations as well as computing the predicted SZ effect signals from numerical simulations.
\end{abstract}


\section{Introduction}
\label{sec:Intro}
%
Free electrons residing inside the deep potential wells of galaxy clusters scatter photons of the cosmic microwave background (CMB), causing a spectral distortion which is commonly referred to as the Sunyaev-Zeldovich (SZ) effect.
The thermal/random motions of electrons in the hot cluster atmospheres lead to the thermal SZ (thSZ) effect \citep{Zeldovich1969}, exhibiting a $y$-type spectral shape related to the up-scattering of CMB photons.
The bulk/directed motion of the electrons, on the other hand, causes a temperature shift in the direction of the cluster, known as the kinematic SZ (kSZ) effect \citep{Sunyaev1980}.

The SZ effect has long been realized as a powerful tool to learn about the formation of structures in the Universe \citep[see][for reviews]{Rephaeli1995ARAA, Birkinshaw1999, Carlstrom2002}. 
Since the effect is independent of redshift, the SZ signal is ideally suited to discover massive galaxy clusters out to high redshifts. 
New generations of ground-based \citep[e.g.,][]{Muchovej2007, Vanderlinde2010, Marriage2011} and space-based \citep{AdeESZCS} CMB experiments have lead to discoveries of several hundred SZ-selected clusters.
Current cluster samples have already provided unique constraints on the dark energy equation of state \citep[e.g.,][]{Wang2004, Gert2006b, Vikhlinin2009, Benson2011, STA11}, the Hubble constant \citep[e.g.,][]{Birkinshaw1991, Hughes1998, Reese2002, Bonamente2006}, and non-Gaussianity in the primordial matter density field \citep{Foley2011}.

Many of the newly discovered SZ-selected clusters contain very hot intracluster plasma  ($k\Te\gtrsim 10\,{\rm keV}$) \citep{Williamson2011, Korngut2011} and exhibit disturbed morphology, indicative of frequent mergers with large peculiar velocity ($v \gtrsim 1000 \, {\rm km s}^{-1}$) \citep{Halverson2009Bullet, Ma2012, Mroczkowski2012}. 
The `Bullet' cluster is one well-studied examples of a high velocity merging system, with $v \simeq 4000 \, {\rm km s}^{-1}$ \citep{Markevitch2002, Markevitch2006, Springel2007, Milosavljevic2007, Mastropietro2008, Lee2010}.
`El Gordo' is another spectacular example of a massive, merging cluster at $z=0.87$ filled with very high temperature gas up to $k\Te= 22\,{\rm keV} \pm 6\,{\rm keV}$ \citep{ElGordo2012}. 
Furthermore, \citet{Mroczkowski2012} recently measured the SZ effect for a triple merger system, finding shock heated gas with temperature $k\Te\geq 20\,{\rm keV}$ and large peculiar motion ($v \simeq 3000 \, {\rm km s}^{-1}$).
Because of the high electron temperature and large kinematic terms, interpretation of the observed SZ signal for such systems clearly must take into account relativistic corrections.
These corrections have been investigated theoretically by several independent groups \citep[e.g.,][]{Wright1979, Rephaeli1995, Challinor1998, Itoh98, Sazonov1998, Nozawa1998SZ, Molnar1999, Ensslin2000, Dolgov2001, Shimon2004, Chluba2005b, Poutanen2010}, and various analytic approximations already provide a good description of the SZ signals in different ranges of temperatures and frequencies.
However, a number of high-resolution SZ experiments, including ALMA\footnote{Atacama Large Millimeter/submillimeter Array}, CARMA\footnote{Combined Array for Research in Millimeter-wave Astronomy}, CCAT\footnote{Cornell Caltech Atacama Telescope}, and MUSTANG\footnote{MUltiplexed Squid TES Array at Ninety GHz}, are underway or planned, promising a dramatic increase in sensitivities, spatial resolution, and spectral coverage over the next few years.
These upcoming SZ experiments should enable a host of new measurements of important cluster properties, including the electron temperature of the intracluster medium (ICM) \citep{Pointecouteau1998, Hansen2002, Zemcov2012}, the peculiar velocity and internal bulk and turbulent gas motions of clusters \citep{Nagai2003, Sunyaev2003, Diego2003}, and non-equilibrium electrons produced by merger and accretion shocks \citep{Markevitch2007, Rudd2009}, to name a few.  
In order to properly interpret these upcoming measurements, it is critical to develop a fast and precise method for computing the SZ effect signal, ideally with accuracy well beyond the precision of future SZ experiments.

In this work, we present a fast and accurate method for computing the relativistic corrections to the SZ effect. 
The approach is based on an alternative derivation of the Compton collision terms, obtained using explicit Lorentz-transformations of the SZ signals from the cluster frame into the observer frame.
This method allows us to clearly identify the source of different kinematic correction terms.
The cluster's peculiar velocity, $\betac = v/c$, makes the electron distribution function effectively anisotropic in the CMB rest frame. 
One way of dealing with the problem is therefore to use the Lorentz-boosted Boltzmann equation in the CMB rest frame \citep[e.g.,][]{Nozawa1998SZ, Challinor1999, Nozawa2006}.
Alternatively, one can treat the problem in the cluster frame. Upon Lorentz-transformation of the isotropic CMB into this frame, the CMB spectrum becomes anisotropic \citep[e.g., see][]{Chluba2011ab}, with motion-induced monopole ($\propto\betacsq$), dipole ($\propto\betac$) and quadrupole ($\propto\betacsq$), due to aberration and Doppler boosting.
These anisotropies are then scattered by hot electrons, imprinting a distortion on the CMB.
Using the Lorentz-invariance of the photon occupation number and line-of-sight number of scatterings one can readily obtain the SZ intensity signal in the observer's frame.
The advantage is that in the cluster frame complications related to kinematic corrections (aberration, retardation and time-dilation effects) and the cluster geometry (see Sect.~\ref{sec:appear}) can be avoided.

The crucial generalization is that we consider not only the scattering of the radiation monopole, but also dipole and quadrupole scattering by hot electrons.
The required kinetic equation for the scattering of anisotropic radiation in lowest order of the electron temperature was recently derived by \citet{Chluba2012}; however, here we include higher order temperature correction to this problem.
We obtain a reformulation of the frequency-dependent basis functions (Sect.~\ref{sec:T_basis_new} and \ref{sec:k_basis_new}) that allow very precise ($\lesssim 0.001\%$  at frequencies $h\nu \lesssim 20 k\Tg$) computation of the SZ signals up to high electron temperature ($k\Te \simeq 25 {\rm keV}$) and large peculiar velocity ($v/c \simeq 0.01$).
Our new approach pushes the precision for the SZ predictions far beyond the sensitivity of future SZ observation, eliminating the need for more expensive numerical integrations of the Boltzmann collision term.
It should therefore be useful when analyzing future high resolution, multi-frequency SZ data for individual clusters or when computing the predicted SZ signals from numerical simulations \citep{Nagai2003, Sunyaev2003, Diego2003, Dolag2005, Battaglia2010}.

With our new derivation, we also show that the kinematic corrections to the SZ {\it intensity} signal given here differ from previously obtained expressions \citep[e.g.,][]{Sazonov1998, Nozawa1998SZ, Nozawa2006, Shimon2004}.
This difference is related to the interpretation of the scattering optical depth integral, which was not explicitly addressed in the earlier calculations (see Sect.~\ref{sec:kSZ_Itoh}).
Although, the differences are small, they could still be relevant for instance when using the SZ effect to confirm the redshift-scaling of the CMB temperature \citep{Battistelli2002, Horellou2005, Luzzi2009, deMartino2012, Avgoustidis2012}, an idea that was suggested long ago \citep{Fabbri1978, Rephaeli1980}.
We also briefly mention additional complications connected with kinematic effects on the measurement that should be taken into account when interpreting future SZ data for individual clusters (see Sect.~\ref{sec:appear}). However, a detailed analysis is beyond the scope of this paper.

One of the main outcomes of this work is {\sc SZpack}\footnote{Available at \url{http://www.Chluba.de/SZpack}}, a numerical library which allows explicit computation of the SZ signals using the full Compton collision integral, but also fast approximation of the numerical result by means of the improved set of frequency-dependent basis functions obtained here. 
This new approach allows calculation of the SZ signals with precision $\lesssim 0.001\%$ over a wide range of parameters at practically no computational cost, overcoming limitations (see Sect.~\ref{sec:thermal_SZ_improved} for more details) of previously derived analytic approximations \citep[e.g.][]{Challinor1998, Itoh98, Sazonov1998} to describe the SZ signals from hot, moving clusters.
With respect to other fast representations of the Boltzmann collision by means of extended tables or fitting functions \citep{Nozawa2000fitting, Itoh2004fittingII, Shimon2004}, our approach provides an interesting alternative, with the derived basis functions being directly informed by the underlying physics of the problem. We furthermore include the kinematic corrections in this representation, describing the SZ intensity signal in the cluster frame and then obtaining the final result by Lorentz-transformation.
This approach also allows us to include the effect of the observer's motion \citep{Chluba2005b} in a simple manner, without extra approximations.

\section{Anisotropic Compton scattering} 
\label{sec:AnisoCS}
In this section we present a brief derivation of the required kinetic equation for Compton
scattering of photons by hot thermal electrons. For now, we neglect the effect of peculiar motion and only consider relativistic correction in different orders of the electron temperature, $\Te$.
However, we describe the scattering of {\it anisotropic} photon distributions in the rest frame of the SZ cluster. This generalization is needed to include kinematic corrections up to second order in the peculiar motion using explicit Lorentz transformations, as recently pointed out by \citet{Chluba2012}.

\subsection{The Boltzmann collision term for a resting cluster}
\label{sec:GenDef}
The time evolution of the photon phase space density $n(t, \nu, \vgh)$
at frequency $\nu$ in the direction\footnote{In the following bold font denotes 3-dimensional 
vectors and an additional hat means that it is normalized to unity.} $\vgh$ under Compton scattering can be described using the
Boltzmann equation \citep[compare with][]{Buchler1976, Itoh98}
\beal
\label{eq:BoltzEq}
\pAb{n(t, \nu, \vgh)}{t}\approx c\int \frac{\id\sigma}{\id \Omega'}\,  \mathcal{F} \id^2 \hat{\gamma}' \id^3 p 
\Abst{,}
\end{align}
where $\id^2 \hat{\gamma}'$ is the solid angle spanned by the incoming photon, $\mathcal{F}$ is the statistical factor, and $\id\sigma/\id \Omega'$ denotes the differential scattering cross section\footnote{More concisely, the factor related to the M{\o}ller relative velocity was absorbed in the definition of $\id\sigma/\id \Omega'$.}.
We assumed that the incoming photon field is spatially homogeneous, which for the problem of interest here is a valid approximation as long as the number of scatterings is small.

To compute the SZ effect one can neglect the effect of electron recoil, since the energy of CMB photons, $h\nu$, is much smaller than the energy of the fast electrons inside the cluster, i.e., $h\nu \ll \gamma \me c^2$.  Here $\gamma=1/\sqrt{1-\beta^2}$ is the Lorentz factor and $\beta$ the dimensionless velocity of the moving electron.
In that limit the Compton scattering cross section reads \citep[e.g., see][]{Jauch1976}
\beal
\label{eq:dsigdO}
\frac{\id\sigma}{\id \Omega'}&\approx\frac{3\,\sigT}{8\pi}\,\left(\frac{\nu'}{\nu}\right)^2 
\frac{1}{\gamma^2\kappa}
\left[1-\frac{\nu'}{\nu}\frac{\alpha_{\rm sc}}{\gamma^2\kappa^2}
+\frac{1}{2}\left(\frac{\nu'}{\nu}\frac{\alpha_{\rm sc}}{\gamma^2\kappa^2}\right)^2
\right],
\end{align}
with $\kappa=1-\beta\mu$, $\alpha_{\rm sc}=1-\mu_{\rm sc}$, where $\mu_{\rm sc}= \vghp\cdot\vgh$ is the cosine of the scattering angle between the incoming and outgoing photon.
Furthermore, $\mu=\vbh\cdot\vgh$ and $\mu'=\vbh\cdot\vghp$ are the direction cosines of the angle between the scattering electron and the incoming and outgoing photon, respectively.
The Thomson scattering cross section is denoted by $\sigT\approx \pot{6.65}{-25}\,{\rm cm^{2}}$. 
The frequency ratio of the incoming and outgoing photon is given by
\beal
\label{eq:energy_shift}
\frac{\nu'}{\nu}&=\frac{1-\beta\mu}{1-\beta\mu'+\frac{h\nu}{\gamma\me c^2 }(1-\mu_{\rm sc})} 
\approx \frac{1-\beta\mu}{1-\beta\mu'}
\Abst{.}
\end{align}
%
Since the temperature of the electron gas obeys $k\Te\ll\me c^2$, 
Fermi-blocking is negligible. In addition stimulated scattering can be omitted so that 
the statistical factor $\mathcal{F}$ may be approximated as
$\mathcal{F}\approx f'\,n' -f\,n$,
where the abbreviations $f=f(\vp),\;f'=f(\vpp)$ for the electron, and
$n=n(\nu, \vgh),\;n'=n(\nu', \vghp)$ for the photon phase space
densities were introduced.
The electron phase space density is isotropic and may be described 
by a relativistic Maxwell-Boltzmann distribution,
\beq\label{eq:relMBD}
f(\vp)
=\frac{\Ne\, e^{-\epsilon(\vp)/\theta_{\rm e}}}{4\pi(\me c)^3 K_2(1/\theta_{\rm e})\,\theta_{\rm e}}
\Abst{,}
\eeq
where $K_2(1/\theta_{\rm e})$ is the modified Bessel function of second kind
with dimensionless electron temperature $\theta_{\rm e}=k\Te/\me c^2$, $\Ne$ is the electron number density, and
$\epsilon(\vp)=\sqrt{1+\eta(\vp)^2}\equiv \gamma$ denotes the
dimensionless energy of the electrons with the dimensionless momentum
$\eta(\vp)=|\vp|/\me c=\gamma \beta$. 
%

With the above definitions, the statistical factor can be cast into the form
$\mathcal{F}\approx f [e^{\Deltaxe}\,n' -n ]$, where we defined $\xe=\frac{h\nu}{k\Te}$ and $\Deltaxe=\xe'-\xe$.
Introducing $\xg=h\nu/k\Tg$ and $\Delta_\nu = (\nu'-\nu)/\nu$, the statistical factor can be rewritten as
\beal
\label{eq:Fappr}
\mathcal{F}/f(\vp)
&\approx 
-n(\xg,\vgh) 
+\sum_{k=0}^\infty  \frac{\Delta^{k}_\nu}{k!}\,\xg^{k}\partial^k_{\xg} n(\xg,\vghp).
\end{align}
where we set $\Deltaxe\simeq (\Tg/\Te) \, \xg \, \Delta_\nu  \approx 0$, neglecting terms that are multiplied by the CMB to electron temperature ratio, $\Tg/\Te$, which for typical clusters is of order $\simeq 10^{-8}$.

We now define the moments of the energy shifts
\beal
\label{eq:Ilmk}
I_{lm}^{k}= \frac{1}{\Ne\,\sigT}\int \frac{\id\sigma}{\id \Omega'}
\, f(\vp)\,\frac{\Delta^{k}_\nu}{k!}\,Y_{lm}(\vghp) \,\id^2 \hat{\gamma}' \id^3 p 
\end{align}
over the scattering cross section. 
Here $Y_{lm}(\vghp)$ denote spherical harmonic functions.
Since $Y_{l(-m)}(\vghp)=(-1)^m\,Y^\ast_{lm}(\vghp)$, it directly follows 
$I_{l(-m)}^{k}\equiv (-1)^m\,(I_{lm}^{k})^\ast$, so that only the integrals 
with $m\geq 0$ ever have to be explicitly computed.
For small $\The$ these moments are all frequency-independent polynomials, which can be explicitly computed up to some order, $k_{\rm max}$, in $\The$.
As we also see below, $I_{lm}^{k}=\lambda^{k}_l \, Y_{lm}(\vgh)$ by symmetry, where $\lambda^{k}_l$ is a temperature-dependent function.
In Appendix~\ref{app:moments} we give the results for the moments of interest up to $\mathcal{O}(\The^{11})$ for monopole scattering, and $\mathcal{O}(\The^{9})$ for dipole and quadrupole scattering
%

%
If we write $n(\xg,\vghp)$ as a spherical harmonic expansion with spherical harmonic coefficients $n_{lm}(\xg)$, then the required Boltzmann equation takes the simple form
\beal
\label{eq:BoltzEqlm_lm}
\pAb{n(\xg,\vgh)}{\tau}
&\approx
-n(\xg,\vgh)
+ 
\sum_{k=0}^{\infty}
\sum_{l,m} 
I_{lm}^{k} \, \xg^{k} \partial^k_{\xg} n_{lm},
\end{align}
where we introduced $\id \tau= c \Ne\,\sigT\id t$. In the optically thin limit, this expression is formally valid for any anisotropic, yet spatially uniform radiation field that is scattered by hot electrons and in which electron recoil and stimulated scatterings can be neglected. 
For the temperature correction to the SZ effect one only needs to consider the scattering of the monopole part of the radiation field, but when considering kinematic corrections, anisotropies in the photon distribution must also be included.

%
With modern computers it is straightforward to carry out the full 5-dimensional collision integral, Eq.~\eqref{eq:BoltzEq}, numerically. 
Here we use the quadrature rules according to \citet{Patterson1968} for each of the independent 1-dimensional integrals. These rules are fully nested and converge rapidly with only a few refinements.
Below we simplify the collision term by reducing the number of integrals and by leveraging symmetries of the problem in different situations. 
To confirm the accuracy of the results we always compare to those obtained with the full 5-dimensional collision term.
With the full collision term it is also possible to compute the SZ signals for more general incoming radiation fields with anisotropies.
The corresponding routines can be found in {\sc SZpack}.

\section{Thermal SZ effect for hot electrons}
\label{sec:thermal_SZ}
To compute the temperature correction to the SZ effect for $\tau\ll 1$ we have to insert $n\approx \nPl=1/(e^{\xg}-1)$ into the Boltzmann equation, Eq.~\eqref{eq:BoltzEqlm_lm}.
Up to $n^{\rm th}$ order in $\The$ we need to compute the derivatives $\xg^{k} \partial^k_{\xg}\nPl$ for $k\leq 2n+2$.
These derivatives can be expressed in several ways, some of which are summarized in Appendix~\ref{app:nPl_derives}.
For explicit computations of the relativistic corrections it is enough to directly use the expressions for $\xg^{k} \partial^k_{\xg}\nPl$ as they are, taking care of the order of different terms to improve the numerical stability (see Appendix~\ref{app:nPl_derives}). 
Together with the moments in Table~\ref{tab:monopole} they determine all correction terms up to the desired order in temperature, in our case $\mathcal{O}(\The^{11})$.
The spectral distortion caused by the thSZ effect in the single scattering approximation is then given by 
\bsub
\label{eq:thSZ_Yk_expression}
\beal
\Delta n_{\rm th}(\xg,\vgh)
&\approx \sum_{k=0}^{10} y^{(k)} Y_k(\xg)
\\
y^{(k)}&= \int \left(\frac{k\Te}{\me c^2}\right)^{k+1} \Ne \sigT c \id t 
\approx \The^{k+1} \Delta \tau \equiv \The^{k} \, y^{(0)},
\label{eq:thSZ_Yk_expression_b}
\end{align}
\esub
where we introduced the generalized $y$-parameter, $y^{(k)}$, and the line-of-sight optical depth\footnote{Unless noted otherwise henceforth we will refer to `optical depth' as quantity that is defined inside the cluster frame. With this convention the optical depth is directly characterizing the number of scatterings along the photon's world line, and hence cleanly separates the effect of scattering and kinematic terms on the SZ signal, as we explain in detail in Sect.~\ref{sec:tau_transform}.}, $\Delta \tau=\int \Ne \sigT c \id t$.
For the second and third equalities of Eq.~\eqref{eq:thSZ_Yk_expression_b}, we also assumed that the cluster is isothermal along the line-of-sight.
Alternatively, one can consider the distortion caused by a single scattering event in one small volume element of the cluster and then sum each contribution along the line-of-sight.
The functions $Y_n$ are determined by 
$Y_n=\sum_{k=1}^{2n+2} a_k^{(n)} \xg^k\partial^k_{\xg} n_{00}(\xg) Y_{00}$ with $n_{00}(\xg) Y_{00} \equiv \nPl(x)$, and the coefficients $a_k^{(n)}$ summarized in Table~\ref{tab:monopole}.

With the definitions given in Appendix~\eqref{app:nPl_derives}, in first order of the electron temperature we have 
\beal
\frac{\Delta n^{(1)}_{\rm th}}{\Delta\tau}
&\approx \The\,\left[4\xg\partial_{\xg} +\xg^2\partial^2_{\xg}\right] \nPl 
= \The\,\mathcal{G}\left[\xg \coth\left(\frac{x}{2}\right)- 4\right]\equiv \The Y_0,
\nonumber
\end{align}
with $\mathcal{G}(x)=x e^x/[e^x-1]^2 = - x \partial_x \nPl(x)$.
This is the well-known result given by \citet{Zeldovich1969} for the thSZ effect, which can be directly derived using the Kompaneets equation \citep{Kompa56, Weymann1965}.
In second order of the electron temperature, we find
\beal
\frac{\Delta n^{(2)}_{\rm th}}{\Delta\tau}
&\approx \The^2
\,\left[10\xg\partial_{\xg} +\frac{47}{2}\xg^2\partial^2_{\xg}
+\frac{42}{5}\xg^3\partial^3_{\xg}+\frac{7}{10}\xg^4\partial^4_{\xg}\right] \nPl 
\equiv \The^2 Y_1.
\nonumber
\end{align}
This expression is equivalent to Eq.~(2.27) in \citet{Itoh98} and the last term in Eq.~(12) of \citet{Sazonov1998}, as can be easily shown using the expressions, Eq.~\eqref{eq:relations_nPl_Itoh}.
We confirmed higher order terms by comparing with the expressions given by \citet{Itoh98}.
Our expressions for the temperature corrections to the thSZ effect alone are also equivalent to those formulae given by \citet{Shimon2004}, although here we obtained two additional orders in the electron temperature.
We also derived temperature corrections up to $20^{\rm th}$ order in electron temperature, but improvements in the convergence were only marginal.

\begin{figure}
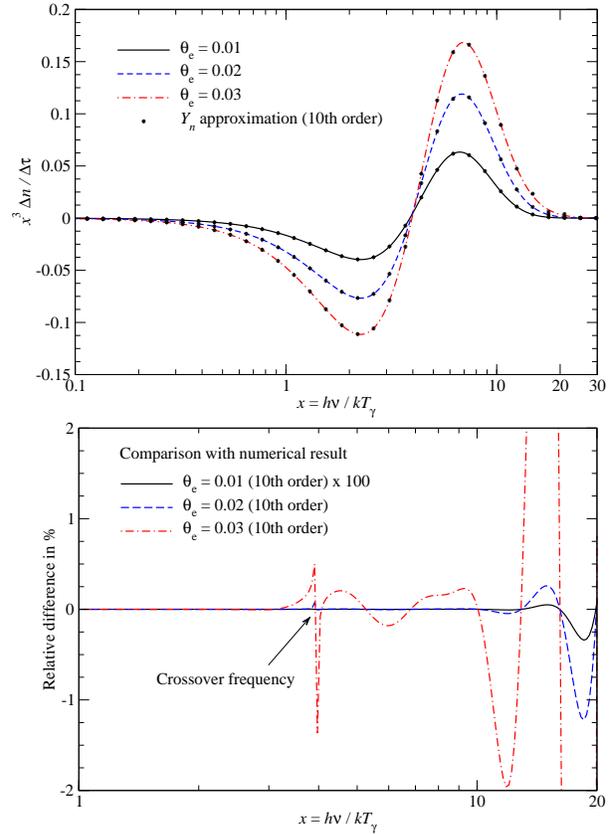

\centering
\includegraphics[width=0.94 \columnwidth]{./eps/DI.Itoh.eps}
\\
\hspace{4mm}\includegraphics[width=0.89 \columnwidth]{./eps/DI.Itoh.diff.eps}
\caption{Thermal SZ effect for an isothermal electron gas with different temperatures, $\The = (k\Te/\me c^2)$. In the upper panel we present the change in the intensity, in dimensionless form, $\Delta I/\Delta \tau\propto x^3 \Delta n/\Delta \tau \equiv x^3 Y(x)$, where $Y(x)=\sum_{k=0}^{10} \The^{k+1}Y_k(x)$, neglecting terms $\mathcal{O}(\The^{12})$. 
The lower panel shows the relative difference with respect to the full numerical result. 
For $\The=0.01$ we multiplied it by 100, demonstrating that the SZ signal is represented to better than $\simeq 0.01\%$ precision over the shown frequency range.
However, for clusters with $k\Te \gtrsim 10\,{\rm keV}$ the Taylor series expansion with the basis functions $Y_k$ does not perform well at $x\gtrsim 10$. Notice that the spike around  the crossover frequency, $x_{\rm cr}\sim 3.83$, is because the SZ signal vanishes there.
The numerical integrals were carried out using {\sc SZpack}.}
\label{fig:DI_Itoh}
\end{figure}
\subsection{Improved representation of the spectral distortions caused by the thermal SZ effect of hot clusters}
\label{sec:thermal_SZ_improved}
Although the moments $I^k_{00}$ have previously been computed to very high orders in temperature, it is known that the associated approximation converges very slowly \citep{Challinor1998, Itoh98, Sazonov1998}.
This is because for high electron temperature ($\gtrsim 10$keV) the width of the scattering kernel \citep[see][for properties of the Compton kernel]{Sazonov2000} becomes sufficiently large that the absolute frequency shift ($\Delta x \simeq x \sqrt{2 \The}\simeq x/5$) caused by Doppler broadening exceeds unity at $x\gtrsim 1/\sqrt{2 \The}\simeq 5$.
Although the relative change in frequency is still small, the simple Taylor series expansion of $\nPl(x')$ around $x$ starts to converge slowly: 
even when taking up to $10^{\rm th}$ order correction in $\The$, the formula provide a rather poor description of the SZ signal at high frequencies\footnote{We found that pushing to much higher order in $\The$ (we computed the $20^{\rm th}$ order) the match degrades significantly.  (cf. Fig.~\ref{fig:DI_Itoh}).}
However, the frequency range $x\simeq 10-20$ is very relevant for interpreting observational data. It is therefore important to compute the thSZ effect from hot clusters precisely even at these high frequencies. Furthermore, higher order temperature corrections are significant near the crossover frequency, $x_{\rm cr}\simeq 3.83$, and improved formulae would be desired.

One straightforward possibility is direct numerical integration of the Boltzmann collision term, Eq.~\eqref{eq:BoltzEq}. Since for the thSZ effect of hot clusters the number of integrals can be analytically reduced to two \citep[e.g.,][]{Wright1979, Ensslin2000, Dolgov2001, Nozawa2009}, the line-of-sight numerical integration for a single temperature is very fast (a few seconds on a standard laptop) and a full analytic description in terms of a Taylor series can be avoided.
However, when analyzing upcoming high resolution SZ data along many lines-of-sight, the inversion of the SZ signal to temperature and electron profile requires a large number of integrals to be taken. In that case it is beneficial to have a simpler and faster method for computing the SZ signal.

One possibility is to produce a simple 2-dimensional table in $(\The, x)$ and then interpolate or fit the results \citep[see e.g.,][]{Nozawa2000fitting, Itoh2004fittingII, Shimon2004}, but for an accurate representation of the collision term ($\lesssim 0.1\%$ say) one needs a rather fine grid, rendering this approach suboptimal. 
In particular, it might be useful to look for a different approach when kinematic corrections are also included.

Here we follow an alternative approach in the middle of the two extremes: one can use a Taylor series in $\The$ for the Boltzmann collision term, but leave $\nPl(x')$ as it is, carrying out the remaining integrals numerically.
This procedure defines a set of frequency-dependent basis functions, $Z_k(x)$, that do not suffer from the same limitation as the $Y_k(x)$ expansion, while still being informed by the underlying physics of the scattering problem.

\subsubsection{Computing the improved basis for the thermal SZ effect}
\label{sec:T_basis_new}
\begin{figure}
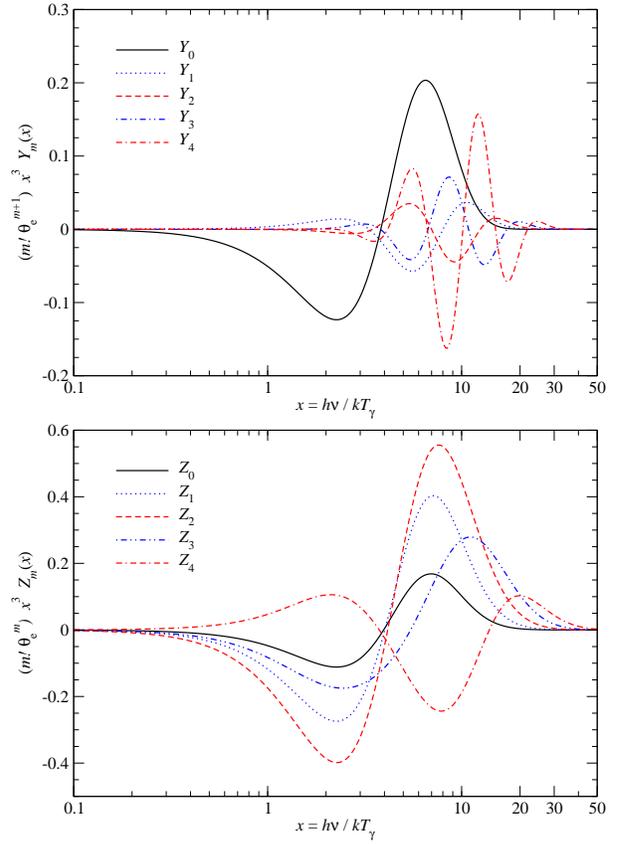

\centering
\includegraphics[width=0.94\columnwidth]{./eps/Y_functions.eps}
\\
\includegraphics[width=0.94\columnwidth]{./eps/Z_functions.eps}
\caption{Comparison of the basis functions $Y_k$ and $Z_k$ for $\theta_{\rm e,0}=0.03$. We scaled both sets by appropriate factors of $k$ and $\The$ to make them comparable in amplitude. One can clearly see that the new set of functions, $Z_k$, exhibits fewer oscillations as $k$ increases. This leads to more stable convergence for higher orders in $\The$. }
\label{fig:Y_Z_functions}
\end{figure}

To compute this new set of basis functions, we first reduce the dimensionality of the collision integral by realizing that for isotropic incoming radiation the azimuthal integrals can be carried out analytically (see Appendix~\ref{app:phi_av_sig} for details). 
The electron temperature enters the problem via the electron distribution function $f(\vp)$, defined in Eq.~\eqref{eq:relMBD}.
If we factorize $f(\vp)$ as 
\beq
\nonumber
f(\vp)
=\left(\frac{\Ne\, e^{-1/\theta_{\rm e}}}{K_2(1/\theta_{\rm e})\,\theta_{\rm e}}\right)
\times \frac{e^{-\Delta\gamma/\theta_{\rm e}}}{4\pi (\me c)^3},
\eeq
with $\Delta \gamma=\gamma-1$, we can define the basis functions\footnote{One could in principle carry out either the integral over $\mu$ or $\mu'$ in addition by defining the logarithmic frequency shift $s=\ln[(1-\beta \mu)/(1-\beta\mu')]$  \citep[cf.][]{Wright1979}, but the resulting expressions are more complicated and numerical integration over 3 dimensions remains sufficiently straightforward with high precision so that we stop here.}
\beal
\label{eq:Zk_def}
Z_k(x)&=   \frac{\mathcal{N}}{k!}\!\!
\int\!\! \frac{\id^2\sigma_0}{\id\mu \id \mu'}
\, \frac{\partial^k e^{-\Delta\gamma/\theta_{\rm e}}}{\partial^k \The}
\left[\nPl(x')-\nPl(x)\right]\!\id\mu \id\mu' \eta^2\!\id \eta
\\[1mm]
\mathcal{N}(\The)
&=\frac{e^{-1/\theta_{\rm e}}}{K_2(1/\theta_{\rm e})\,\theta_{\rm e}}\approx 
\frac{4\pi}{(2\pi\The)^{3/2}}\left[1-\frac{15}{2}\The+\frac{345}{128}\The^2+\mathcal{O}(\The^3)\right]
\nonumber
\end{align}
which can be computed numerically for some fixed temperature, $\The\equiv \theta_{\rm e, 0}$. 
The azimuthally averaged cross section $\id^2 \sigma_0/\id \mu\id \mu'$ is given by Eq.~\eqref{eq:sig_0}.
Then the distortion caused by the thSZ effect of a hot cluster is determined by
\beal
\label{eq:thSZ_Zk_expression}
\Delta n_{\rm th}(\xg,\vgh)
&\approx \sum_{k=0}^{k_{\rm max}} \Delta z^{(k)} Z_k(\xg),
\\
\Delta z^{(k)}&\equiv  \int \frac{\mathcal{N}(\The)}{\mathcal{N}(\theta_{\rm e, 0})} \left(\The-\theta_{\rm e, 0}\right)^{k}\! \id \tau 
\nonumber
\approx 
\frac{\mathcal{N}(\The)}{\mathcal{N}(\theta_{\rm e, 0})} \left(\The-\theta_{\rm e, 0}\right)^{k} \Delta \tau,
\end{align}
where in the last step we assumed that the electron gas is isothermal over the distance $\id l= c\id t$.
This assumption is always justified when choosing $\id l$ sufficiently small, i.e., by considering the scattering effect for a small volume element of the cluster.
The derivatives $\partial^k_{\The} e^{-\Delta \gamma/\The}$ can be performed analytically for any $k$.
In Fig.~\ref{fig:Y_Z_functions} we compare the first few functions $Z_k$ with $Y_k$. One can see that the new set of functions, $Z_k$, exhibits fewer oscillations as $k$ increases. This leads to more stable convergence for higher orders in $\The$.
%

\begin{figure}
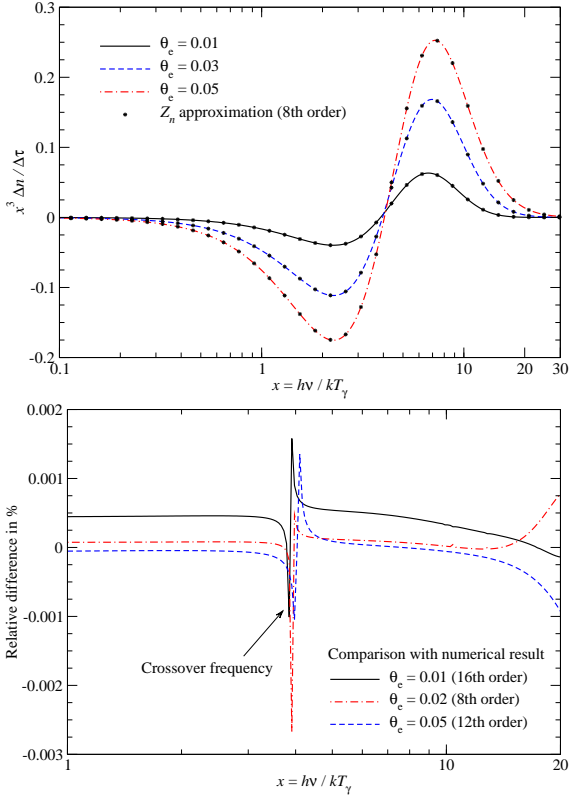

\centering
\includegraphics[width=0.89 \columnwidth]{./eps/DI.Z.eps}
\\
\includegraphics[width=0.89 \columnwidth]{./eps/DI.Z.diff.eps}
\caption{Same as Fig.~\ref{fig:DI_Itoh} but for slightly higher electron temperature and using the expansion, Eq.~\eqref{eq:thSZ_Zk_expression}, for different orders. Clearly the convergence is excellent at all frequencies of interest.}
\label{fig:DI_Z}
\end{figure}

In Fig.~\ref{fig:DI_Z} we illustrate the performance of the new expansion. We computed the basis functions, $Z_k$, at reference temperature $\theta_{\rm e, 0}=0.03$ up to the $20^{\rm th}$ derivative in $\The$ using {\sc SZpack}. Clearly the full numerical result is represented very accurately (with precision comparable to $0.001\%$) over a wide range of frequencies and temperatures. We found that the expansion works well even at temperatures of $\The > 0.05$. At low temperatures, $\The \lesssim 0.01$, it is better to rely on the asymptotic expansion based on $Y_k$, since errors start to become larger. However, the combined description with $Y_k$ and the new functions $Z_k$ allows very precise computation of the thSZ effect in an extremely efficient way. A combination of both approximations is implemented in {\sc SZpack}.
Alternatively, one could produce a set of basis functions $Z_k$ around different reference temperatures and in this way compress the representation of the full integral in an economic way, while achieving even higher precision at the same time. This is especially important when considering effects in second order of the cluster's peculiar motion (velocity of $\simeq 3000{\rm km/s}$ say) for which relative precision better than $\simeq 10^{-4}$ is required.

\section{Kinematic corrections to the SZ signal}
\label{sec:kinematic}
%
To derive the modifications caused by the motion of the cluster with respect to the CMB and the observer, we describe the scattering process in the rest frame of the cluster, where the incoming radiation field is anisotropic due to kinematic effects.
We can use the formalism developed in Sect.~\ref{sec:AnisoCS} to compute the introduced distortion by thinking of the problem as scattering of monopolar through quadrupolar part of the photon distribution.
Transformation of the SZ signal into the observer frame can then be achieved by simple Lorentz transformation (Sect.~\ref{sec:beta_problem} and \ref{sec:trans_general_obs}), without additional loss of precision.
We restrict the main derivation to kinematic corrections up to second order in the clusters velocity, $\vbetac$, but neglect the effect caused by terms in $\mathcal{O}(\betac^3)$ and scattering of primordial large-scale CMB anisotropies.
We briefly discuss these effects in Sect.~\ref{sec:small_corrs}, arguing that close to the crossover frequency these only contribute at the level of $0.1\%$ relative to the kSZ signal.
Inside the cluster frame, the computation of the SZ signal is straightforward, and complications related to aberration, retardation and time-dilation effects, as well as geometrical aspects can be avoided. These, however, enter the problem when transforming to a general observer, as we discuss in Sect.~\ref{sec:appear}.

\subsection{SZ signal in the moving cluster frame}
As a first step, we transform the photon distribution from the CMB rest frame into the cluster frame which is moving with a peculiar velocity $\vbetac$. Aligning the $z$-axis with the direction of the cluster's motion, it is straightforward to show that 
\citep[cf.,][]{Chluba2012}
\beal
\label{eq:n_Lorentz_examples}
n^{\rm c}_{00}(\xgc)&\approx n_{00}
+\frac{\betacsq}{2}\left[\xgc \partial_{\xgc}+\frac{1}{3}\xgc^2 \partial^2_{\xgc}\right] n_{00},
\nonumber\\
n^{\rm c}_{10}(\xgc)
&\approx 
\alpha_1\,\betac  \, \xgc \partial_{\xgc} n_{00} 
\nonumber\\[1mm]
n^{\rm c}_{20}(\xgc)
&\approx 
\alpha_2\,\betacsq\,\xgc^2 \partial^2_{\xgc}n_{00}
\end{align}
in the cluster frame. In the chosen coordinates the photon distribution is azimuthally symmetric about $\vbetach$, with all other spherical harmonic coefficients vanishing in $\mathcal{O}(\betac^2)$.
We neglected any intrinsic large-scale anisotropies of the CMB and set $n_{00}=\sqrt{4\pi} \,\nPl(\xgc)$.
Also, $\alpha_1=1/\sqrt{3}$ and $\alpha_2=1/(3\sqrt{5})$, and $\xgc=h \nu_{\rm c} / k\Tg$, where $\nu_{\rm c}$ is the photon frequency evaluated in the cluster frame and $\Tg$ defines the CMB monopole temperature in the CMB rest frame\footnote{The value of $\Tg$ can be determined, a point that we clarify below.}.

As Eq.~\eqref{eq:n_Lorentz_examples} shows, in first order of $\betac$ a dipolar anisotropy
is induced by the cluster's motion relative to the CMB rest frame.
The spectrum for this part of the photon distribution corresponds to a temperature shift, $\mathcal{G}(\xgc)=-\xgc \partial_{\xgc} \nPl =\xgc e^{\xgc}/(e^{\xgc}-1)^2$.
In second order of $\betac$ a quadrupolar anisotropy and a small correction to the monopole appears, where both Doppler boosting and the aberration effects contribute \citep{Challinor2002, Kosowsky2010, Amendola2010, Chluba2011ab, Chluba2012}.
The spectrum of the quadrupolar part is $\mathcal{Q}(\xgc)=\xgc^2 \partial^2_{\xgc} \nPl = Y_0(\xgc) + 4 \mathcal{G}(\xgc)$, which has a $y$-type dependence. Similarly, the correction to the monopole exhibits a $y$-type distortion.
Higher order correction terms (i.e., the motion-induced octupole, $\propto \betac^3$, and hexa-decupole, $\propto \betac^4$) can be easily obtained, but for typical peculiar velocities ($\betac\simeq 10^{-3}-10^{-2}$) these are very small (see Sect.~\ref{sec:small_corrs}).
Also, the associated Taylor series in $\betac$ converges quickly, which is in stark contrast to the expansion in small electron temperature and frequency shift discussed above, which only converges asymptotically.

Inserting Eq.~\eqref{eq:n_Lorentz_examples} into Eq.~\eqref{eq:BoltzEqlm_lm} and collecting terms we have
\beal
\label{eq:BoltzEqlm_lm_cluster}
\pAb{n^{\rm c}(\xgc,\vgh^{\rm c})}{\tau^{\rm c}}
&\approx
\sum_{k=1}^{\infty}
\hat{I}_{00}^{k} \, \xgc^{k} \partial^k_{\xgc} 
\left[
1+\frac{\betacsq}{2}\left(\xgc \partial_{\xgc}+\frac{1}{3}\xgc^2 \partial^2_{\xgc}\right)
\right]
\nPl
\nonumber\\
&
+\betac \muc^{\rm c}
\left[
\Delta \hat{I}_{10}^{0}\xgc \partial_{\xgc} \nPl
+ 
\sum_{k=1}^{\infty}
\hat{I}_{10}^{k} \, \xgc^{k} \partial^k_{\xgc} \xgc \partial_{\xgc} \nPl
\right]
\\
&
+\frac{1}{3}\,\betacsq P_2(\muc^{\rm c})
\left[
\Delta \hat{I}_{20}^{0}\xgc^2 \partial^2_{\xgc} \nPl
+ 
\sum_{k=1}^{\infty}
\hat{I}_{20}^{k} \, \xgc^{k} \partial^k_{\xgc} \xgc^2 \partial^2_{\xgc} \nPl
\right].
\nonumber
\end{align}
Here $P_2(x)=[3x^2-1]/2$ is the second Legendre polynomial and $\muc^{\rm c}$ is the direction cosine of the angle between $\vbetac$ and the outgoing photon, $\vgh^{\rm c}$, evaluated in the cluster frame.
Furthermore, we defined $\Delta \hat{I}_{l0}^{0}=\hat{I}_{l0}^{0}-1$, and the moments $\hat{I}_{lm}^{k}={I}_{l0}^{k}/Y_{lm}(\vgh^{\rm c})$, which are only functions of the electron temperature, $\The^{\rm c}=k\Te^{\rm c}/\me c^2$, fixed in the cluster frame.
Also, $\id \tau^{\rm c}\equiv \Ne^{\rm c}\sigT c \id t^{\rm c}$ defines the Thomson optical depth, with electron number density, $\Ne^{\rm c}$, again in the cluster frame.

Equation~\eqref{eq:BoltzEqlm_lm_cluster} is still the general expression, describing the scattering of CMB photons in the cluster frame, written in a compact way. The required moments are summarized in Appendix~\ref{app:moments}. 
Any observer that moves with a velocity $\vbetac$ relative to the CMB measures a spectral distortion in the direction of the cluster that is given by this expression.
By including as many moments, $\hat{I}_{l0}^{k}$, as is required to converge for a given cluster temperature, this is formally the exact result, provided that the value for the peculiar motion of the cluster allows truncation of the series in $\betac$ at second order.
If higher orders in $\betac$ are needed then additional moments $\hat{I}_{l0}^{k}$ with $l>2$ must be computed. 
However, the series in $\betac$ converges rapidly, so that Eq.~\eqref{eq:BoltzEqlm_lm_cluster} should be enough.
As expected, when truncating the moment-hierarchy at some finite temperature, we again encounter problems of convergence for hot clusters.
We will show how to overcome these limitations below.

We also mention that, as expected, the SZ signal is independent of the azimuthal angle, $\phi_{\rm c}$, of $\vbetac$ with respect to the line-of-sight, reflecting the symmetry of the scattering process. For the distortion only the projections of the anisotropic photon distribution onto $Y_{l0}$ matters, once the $z$-axis is re-aligned with the direction of $\vgh^{\rm c}$.
This statement is no longer true when the polarization signature caused by the kSZ effect is considered \citep{Sunyaev1980, Sazonov1999}; in this case $\phi_{\rm c}$ defines the direction perpendicular to the linear polarization plane.
Internal motions of the ICM can cause interesting effects in the polarization signal, in principle allowing to study the transverse intra-cluster velocity field \citep{Chluba2002, Diego2003}.
However, a treatment of the polarization effects, including temperature corrections, is beyond the scope of this paper.
%

\begin{figure}
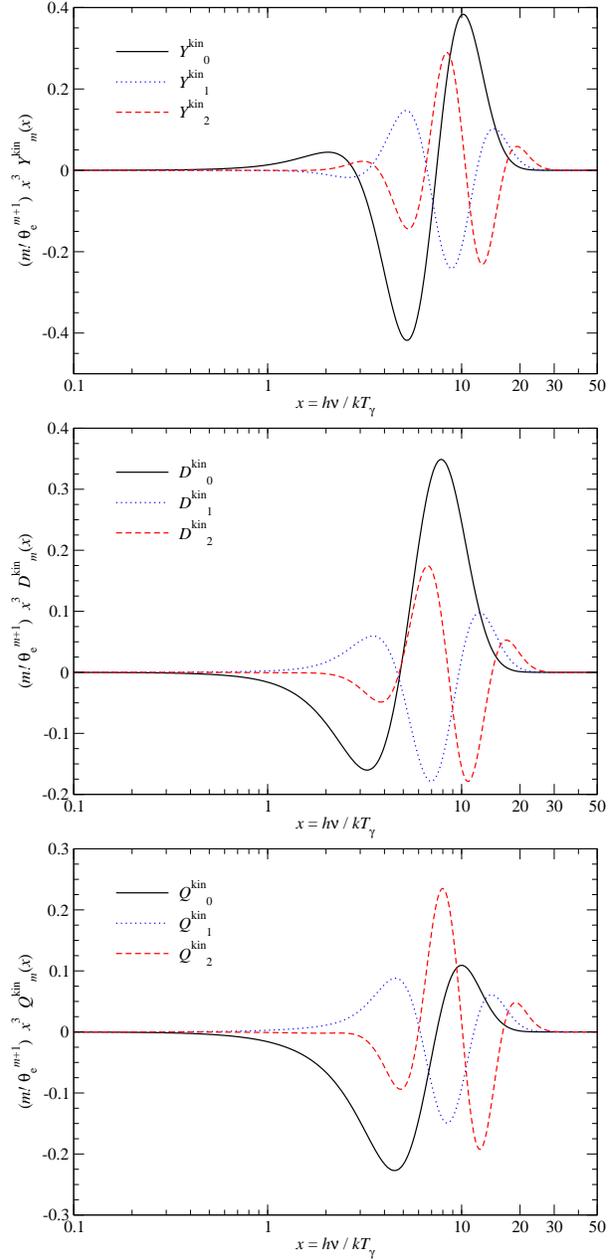

\centering
\includegraphics[width=0.94\columnwidth]{./eps/Ykin_functions.eps}
\\
\includegraphics[width=0.94\columnwidth]{./eps/Dkin_functions.eps}
\\
\includegraphics[width=0.94\columnwidth]{./eps/Qkin_functions.eps}
\caption{First few functions $Y^{\rm kin}_k$, $D^{\rm kin}_k$ and $Q^{\rm kin}_k$. We rescaled by appropriate factors of $k$ and $\The=0.03$ to make them comparable in amplitude.}
\label{fig:Ykin_functions}
\label{fig:Dkin_Qkin_functions}
\end{figure}

\subsubsection{Effect on the monopole spectrum inside the cluster frame}
\label{sec:monopole_scattering_sect}
The first sum of Eq.~\eqref{eq:BoltzEqlm_lm_cluster} describes the effect of scattering on the monopole seen in the cluster frame, with second order kinematic corrections to the Planckian spectrum included.
The first contribution to this sum is the thSZ effect, which affects the purely Planckian part as discussed in the previous section.
The other two terms $\propto \betac^2$ can be included as corrections to the functions, $Y_k$. 
Since we have a simple and numerically stable method for computing derivatives of the Planck spectrum (see Appendix~\ref{app:nPl_derives_closed}), we can obtain the associated spectral function, $Y^{\rm kin}_k(x)$ using the moments summarized in Table~\ref{tab:monopole}. 
With the simple operator relations
\beal
\label{eq:op_rel}
x^{k} \partial^k_{x} x \partial_{x}&=k x^{k} \partial^k_{x}+x^{k+1} \partial^{k+1}_{x}
\\
x^{k} \partial^k_{x} x^2 \partial^2_{x}&=k(k-1) x^{k} \partial^k_{x}+2k x^{k+1} \partial^{k+1}_{x}
+x^{k+2} \partial^{k+2}_{x},
\nonumber
\end{align}
the effect of scattering on the monopole part of the photon distribution, $n^{\rm c}_{0}=n^{\rm c}_{00}(\xgc) Y_{00}$, is given by
\beal
\label{eq:Y_kin}
\pAb{n^{\rm c}_{0}(\xgc,\vgh^{\rm c})}{\tau^{\rm c}}
&\approx \The^{\rm c} \sum_{k=0}^{10} (\The^{\rm c})^k \left[ Y_k(\xgc)+\betac^2 Y^{\rm kin}_k(\xgc)\right]
\\ 
&\!\!\!\!\!\!\!\!\!\!\!\!\!\!\!\!\!\!\!\!\!\!\!\!\!\!\!
Y^{\rm kin}_n=\frac{1}{6}\,\sum_{k=1}^{2n+2} a_k^{(n)} 
\left[ k(k+2) \xgc^k\partial^k_{\xgc}  
+ (2k+3) \xgc^{k+1}\partial^{k+1}_{\xgc} \!\!
+\xgc^{k+2}\partial^{k+2}_{\xgc}\right]\nPl.
\nonumber
\end{align}
The coefficients $a_k^{(n)}$ are given in Table~\ref{tab:monopole}.
This expression defines the correction of the SZ signal in second order of $\betac$ arising from the scattering of the monopole part of the photon distribution in the cluster rest frame.
With the definitions of Appendix~\ref{app:nPl_derives_Itoh_GX} we can, for example, write
\beal
\label{eq:Y_kin_examples}
Y^{\rm kin}_0=\frac{50}{3}\mathcal{G}+\!\frac{14}{3}Y_0
-\!\frac{11}{6}\xgc\mathcal{G}(6\mathcal{G}+\xgc)
+\frac{1}{6} \xgc\mathcal{G}\tilde{X}(12\mathcal{G}+\xgc)
\end{align}
for the lowest order term in electron temperature. 
Here we introduced $\tilde{X}= x \coth\left(\frac{x}{2}\right)$.
The first few functions $Y^{\rm kin}_k$ are shown in Fig.~\ref{fig:Ykin_functions}.
As for the $Y_k$-basis of the thSZ effect we can observe strong oscillations for larger values of $k$, indicating that the associated series in $\The$ converges slowly.
The correction from the motion-induced monopole is the smallest among all kinematic corrections to the SZ signal, as we discuss below.

\subsubsection{Effect on the dipole and quadrupole part of the spectrum inside the cluster frame}
The next two terms in Eq.~\eqref{eq:BoltzEqlm_lm_cluster} are caused by scattering of the dipolar and quadrupolar anisotropy in the cluster frame.
The contributions $\propto \Delta \hat{I}_{l0}^{0}$ include the effect of Thomson scattering (no energy exchange between electrons and photons) and temperature corrections to the Compton scattering cross section.
Comparing with Eq.~\eqref{eq:n_Lorentz_examples} it is clear that these terms leave the spectrum of the incoming radiation unchanged, but only modify their amplitude by scattering photons in and out of the line-of-sight. 
According to the moments given in the Tables~\ref{tab:dipole} and \ref{tab:quadrupole}, in second order of the electron temperature we have
$\Delta \hat{I}_{10}^{0}\approx - 1 -\frac{2}{5}\The^{\rm c} -\frac{1}{5} (\The^{\rm c})^2$ and $\Delta \hat{I}_{20}^{0}\approx -9/10 -\frac{3}{5}\The^{\rm c} +\frac{183}{70} (\The^{\rm c})^2$.
An observer in the rest frame of the moving cluster sees a dipole and quadrupole spectrum with $n^{\rm c}_{10}(\xgc)$ and $n^{\rm c}_{20}(\xgc)$ given by Eq.~\eqref{eq:n_Lorentz_examples}.
When looking towards the cluster the photon fluxes in the dipolar and quadrupolar parts of the photon distribution are reduced in both cases, since scattering out of the line-of-sight dominates. 
For the dipolar anisotropy in the cluster frame the leading order term describes the kSZ, $\Delta n=\Delta \tau^{\rm c}\betac\muc^{\rm c} \mathcal{G}(\xgc)$ \citep{Sunyaev1980}, while for the quadrupole scattering we have the quadratic kSZ effect (qkSZ), $\Delta n=-\frac{3}{10} \Delta \tau^{\rm c}\betac^2 P_2(\muc^{\rm c}) \xgc \mathcal{G}(\xgc)\coth(\xgc/2)=-\frac{3}{10} \Delta \tau^{\rm c}\betac^2 P_2(\muc^{\rm c}) \mathcal{Q}(\xgc)$.
Here we introduced $\mathcal{Q}(x)=4\mathcal{G}(x)+Y_0(x)=\mathcal{G}(x) \tilde{X}(x)$.

The other terms in Eq.~\eqref{eq:BoltzEqlm_lm_cluster} describe changes in the spectrum of the dipole and quadrupole caused by the scattering process, with energy transfer from the electrons to the photons.
For the total effect on the dipole part of the photon distribution we therefore have
\beal
\label{eq:D_kin}
\pAb{n^{\rm c}_{1}(\xgc,\vgh^{\rm c})}{\tau^{\rm c}}
&\approx
\betac\muc^{\rm c}\left[\mathcal{G}(\xgc)
+ \The^{\rm c} \sum_{k=0}^{8} (\The^{\rm c})^k D^{\rm kin}_k(\xgc)
 \right]
\\ 
&\!\!\!\!\!\!\!\!\!\!\!\!\!\!\!\!\!\!\!\!\!\!\!\!\!\!\!
D^{\rm kin}_n=\sum_{k=0}^{2n+2} d_k^{(n)} 
\left[ k \xgc^k\partial^k_{\xgc}  
+  \xgc^{k+1}\partial^{k+1}_{\xgc} \right]\nPl,
\nonumber
\end{align}
where we used the operator relations, Eq.~\eqref{eq:op_rel}, and $d_k^{(n)}$ according to Table~\ref{tab:dipole}.
Similarly, for the effect on the quadrupole we find
\beal
\label{eq:Q_kin}
\pAb{n^{\rm c}_{2}(\xgc,\vgh^{\rm c})}{\tau^{\rm c}}
&\approx
\betacsq P_2(\muc^{\rm c})\left[
-\frac{3}{10} \mathcal{Q}(\xgc)
+ \The^{\rm c} \sum_{k=0}^{8} (\The^{\rm c})^k Q^{\rm kin}_k(\xgc)
 \right]
\nonumber\\ 
&\!\!\!\!\!\!\!\!\!\!\!\!\!\!\!\!\!\!\!\!\!\!\!\!\!\!\!
Q^{\rm kin}_n=\frac{1}{3}\sum_{k=0}^{2n+2} q_k^{(n)} 
\left[ 
k(k-1) \xgc^k\partial^k_{\xgc}  
+ 2k \xgc^{k+1}\partial^{k+1}_{\xgc} 
+ \xgc^{k+2}\partial^{k+2}_{\xgc} \right]\nPl,
\end{align}
with $q_k^{(n)}$ defined according to Table~\ref{tab:quadrupole}.
The first few functions $D^{\rm kin}_k$ and $Q^{\rm kin}_k$ are shown in Fig.~\ref{fig:Dkin_Qkin_functions}.
We also have
\beal
\label{eq:DQ_kin_examples}
D^{\rm kin}_0(\xgc) &=-\frac{38}{5}\mathcal{G}-\frac{12}{5}Y_0
+\frac{2}{5}\xgc\mathcal{G}(6\mathcal{G}+\xgc)
\\
\nonumber
Q^{\rm kin}_0(\xgc)&=\frac{8}{15}\mathcal{G}+\frac{2}{15}Y_0
-\frac{4}{15}\xgc\mathcal{G}(6\mathcal{G}+\xgc)
+\frac{1}{30}\xgc\mathcal{G}\tilde{X}(12\mathcal{G}+\xgc)
\end{align}
in lowest order term of the electron temperature.

Among all kinematic corrections to the SZ signal, the kSZ effect in first order of the cluster velocity is the largest. Temperature corrections to the kSZ can reach $\simeq 10\%-20\%$ at high frequencies and are therefore larger than the correction from the qkSZ effect.
Temperature corrections to the qkSZ effect are already very small, but still larger or comparable to the signal related to the motion-induced monopole (Sect.~\ref{sec:monopole_scattering_sect}). However, as we show below, it is easy to include all these effects using an approach similar to the thSZ effect case, at no additional computational cost.
%

\begin{figure}
\centering
\includegraphics[width=0.94\columnwidth]{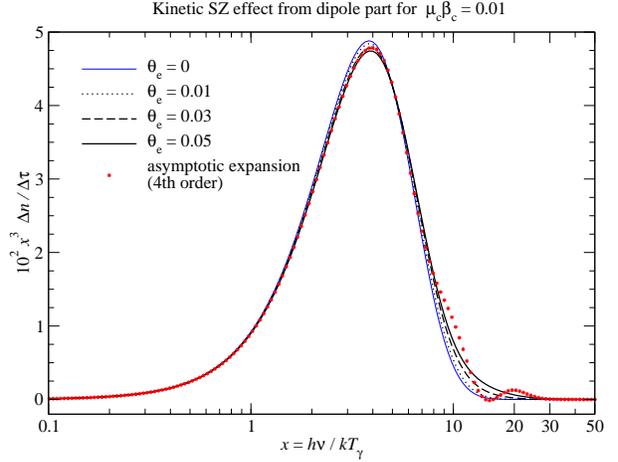}
\caption{Kinematic correction to the SZ signal caused by the motion-induced dipole ($\propto\betac$) in the cluster frame for different temperatures. In the limit $\The=0$ we have the usual kSZ effect. For $\The=0.05$ we also show the result of the asymptotic expansion, Eq.~\eqref{eq:D_kin}, with $4^{\rm th}$ order temperature corrections included, indicating slow convergence at high frequencies.}
\label{fig:kSZ_T}
\end{figure}

\begin{figure}
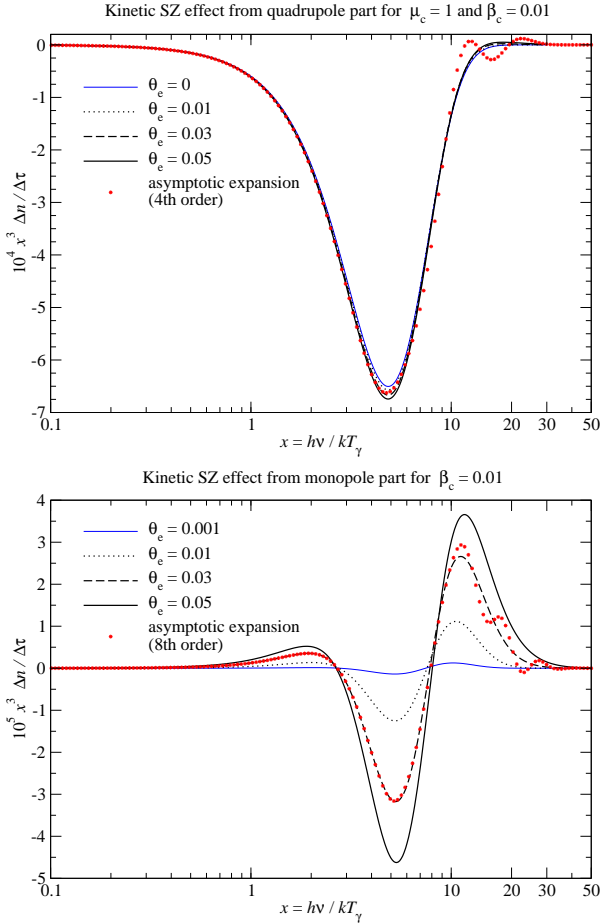

\centering
\includegraphics[width=0.94\columnwidth]{./eps/DI_kSZ.Q.b_0.01.var_T.eps}
\\[1mm]
\includegraphics[width=0.94\columnwidth]{./eps/DI_kSZ.M.b_0.01.var_T.eps}
\caption{Kinematic correction to the SZ signal in the cluster frame to second order of $\betac$ for different temperatures.
The upper panel shows the effect related to the motion-induced quadrupole. In the limit $\The=0$ we obtain the qkSZ effect. For $\The=0.05$ the result of the asymptotic expansion, Eq.~\eqref{eq:Q_kin}, with $4^{\rm th}$ order temperature corrections included is also shown.
---
The lower panel shows the effect related to the motion-induced monopole. 
It vanishes in the limit $\The=0$. For $\The=0.03$ we also give the result of the asymptotic expansion, Eq.~\eqref{eq:Q_kin}, with $8^{\rm th}$ order temperature corrections included, indicating slow convergence at high frequencies.
}
\label{fig:kSZ_T_Q}
\end{figure}

\subsubsection{Comparing with numerical results}
\label{sec:cluster_frame_dist}
It is straightforward to numerically compute the spectral distortion in the rest frame of the cluster. 
By inserting the transformed occupation number, $n^{\rm c}(\xgc, \vgh^{\rm c})=\nPl(\gamma_{\rm c} \xgc(1+\vbetac\cdot\vgh^{\rm c}))$, into the Boltzmann equation, Eq.~\eqref{eq:BoltzEq}, one can directly integrate the collision term to all orders in $\betac$ and $\The$.
We implemented this case as one of the routines in {\sc SZpack} to confirm the results obtained with the simplified approach discussed below.
This also allows us to include the scattering of primordial CMB anisotropies in the cluster rest frame, but both higher order terms in $\betac$ and the scattering of CMB anisotropies cause very small additional signals, which are only noticeable close to the crossover frequency, where the thSZ effect vanishes (see Sect.~\ref{sec:small_corrs}).

\begin{figure}
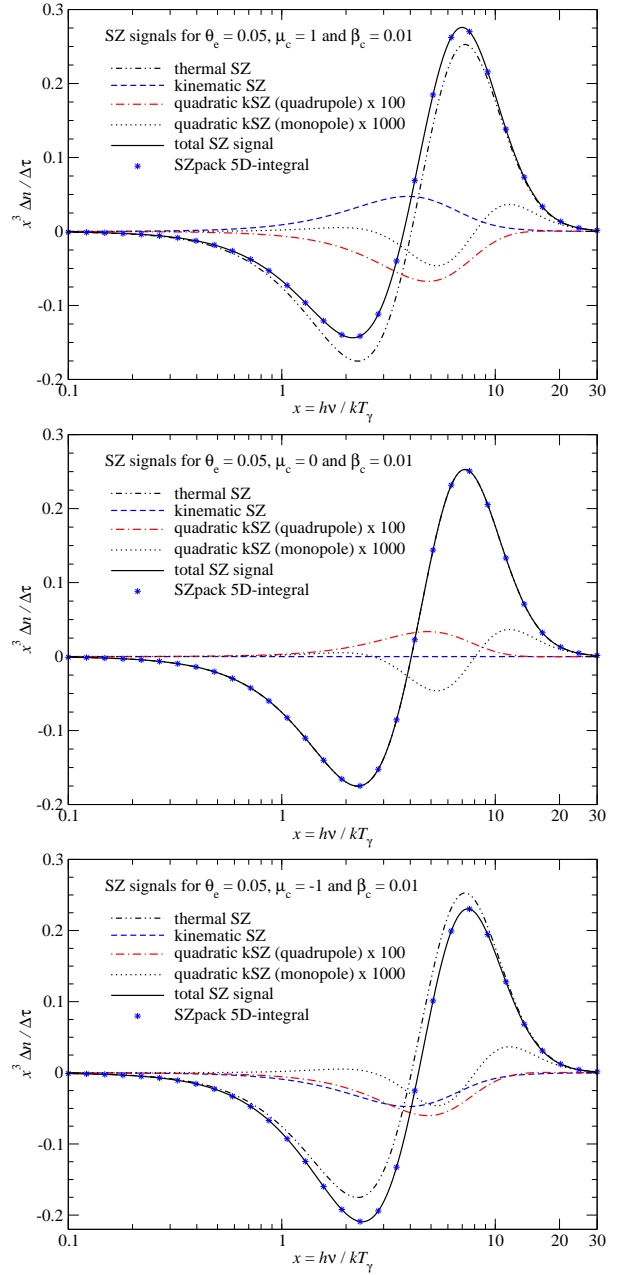

\centering
\includegraphics[width=0.94\columnwidth]{./eps/DI_SZ_signals.The_0.05.betac_0.01.eps}
\\
\includegraphics[width=0.94\columnwidth]{./eps/DI_SZ_signals.The_0.05.betac_0.01.mu_0.eps}
\\
\includegraphics[width=0.94\columnwidth]{./eps/DI_SZ_signals.The_0.05.betac_0.01.mu_m1.eps}
\caption{Comparison of the different SZ signals for $\The=0.05$, $\betac=0.01$ and $\muc=1, 0, -1$ from top to bottom panel. All temperature corrections are included for each of the contributions, and the curves were obtained using the 3-dimensional integrals for monopole through quadrupole scattering.
We also show the result obtained with {\sc SZpack} carrying out the 5-dimensional collision integral explicitly.
Notice that we multiplied the signals in second order of $\betac$ by appropriate factors to make them more visible.}
\label{fig:SZ_signals_comp}
\end{figure}

%
We start by considering the effect of scattering on the dipolar anisotropy in the cluster frame.
We can ease the numerical integration by using the symmetries of the scattering process: with respect to the outgoing photon direction $\vgh^{\rm c}$ we only have to consider the azimuthally symmetric part of the photon distribution function, $n_1(x',\vghp)=\betac\muc^{\rm c} x'\partial_{x'} \nPl(x')\,\mu_{\rm sc}$, with $\mu_{\rm sc}= \vghp \cdot \vgh^{\rm c}$, since the other contributions all average out.
For the dipolar part of the photon distribution we therefore have
\beal
\label{eq:BoltzEq_dipole}
\pAb{n^{\rm c}_1(\xgc, \vgh^{\rm c})}{\tau^{\rm c}}&= 
\betac\muc^{\rm c}\left[
\mathcal{G}(\xgc)+C^{\rm kin}(\xgc, \The^{\rm c})\right]
\nonumber
\\
C^{\rm kin}(\xgc, \The^{\rm c})&=
-\mathcal{N}\!\!
\int\!\! \frac{\id^2\sigma_1}{\id\mu \id \mu'}
\, e^{-\Delta\gamma/\theta^{\rm c}_{\rm e}}\,\mathcal{G}(x_{\rm c}') \id\mu \id\mu' \eta^2\!\id \eta,
\end{align}
where $\id^2 \sigma_1/\id \mu\id \mu'$ is given by Eq.~\eqref{eq:sig_1}.
The dependence on the electron temperature arises from the integral $C^{\rm kin}(\xgc, \The^{\rm c})$, which accounts for additional redistribution of photons (up-scattering) over frequency by energy exchange with the thermal electrons.

The results of the numerical integration are shown in Fig.~\ref{fig:kSZ_T} for different electron temperatures. The case $\The=0$ corresponds to the normal kSZ effect.
Temperature corrections are visible only at high frequencies around $x\simeq 1$ and $10$, reaching $\simeq 10\%-20\%$ relative to the kSZ signal. 
For $\The=0.05$ we also show the result of the asymptotic expansion, Eq.~\eqref{eq:D_kin}, with $4^{\rm th}$ order temperature corrections included. At higher frequencies convergence gets slower and including more orders did not further improve the agreement with the full numerical result.
On the other hand, for low temperatures ($\The\simeq 0.01$) we found good agreement of the expansion, Eq.~\eqref{eq:D_kin}, with the full numerical result.
We also confirmed numerically that the full 5-dimensional collision integral for the dipole part is correctly represented by Eq.~\eqref{eq:BoltzEq_dipole}.

The corrections in second order of $\betac$ can be obtained in a similar manner. For the motion-induced distortion of the monopole, defining $\mathcal{M}(x)=\mathcal{G}(x)+Y_0(x)$, we find
\beal
\label{eq:BoltzEq_monopole_corr}
\pAb{\Delta n^{\rm c}_0(\xgc, \vgh^{\rm c})}{\tau^{\rm c}}
&= 
\betacsq \, Z^{\rm kin}(\xgc, \The^{\rm c})
\\[1mm]
Z^{\rm kin}(\xgc, \The^{\rm c})
&=
\frac{\mathcal{N}}{6}\!\!
\int\!\! \frac{\id^2\sigma_0}{\id\mu \id \mu'}
\, e^{-\Delta\gamma/\theta^{\rm c}_{\rm e}}
\left[\mathcal{M}(\xgc')-\mathcal{M}(\xgc)\right]\!\id\mu \id\mu' \eta ^2\!\id \eta
\nonumber
\end{align}
using Eq.~\eqref{eq:sig_0} and \eqref{eq:Zk_def}. Notice that $Z^{\rm kin}(\xgc, \The^{\rm c})=0$ for $\The^{\rm c}\rightarrow 0$.
For the quadrupole terms we have
\beal
\label{eq:BoltzEq_quadrupole}
\pAb{n^{\rm c}_2(\xgc, \vgh^{\rm c})}{\tau^{\rm c}}&= 
\betacsq P_2(\muc^{\rm c})\left[
-\frac{3}{10}\mathcal{Q}(\xgc)+S^{\rm kin}(\xgc, \The^{\rm c})\right]
\\[1mm]
S^{\rm kin}(\xgc, \The^{\rm c})&=
\frac{\mathcal{N}}{3}\!\!
\int\!\! 
\frac{\id^2\sigma_2}{\id\mu \id \mu'}
\, e^{-\Delta\gamma/\theta^{\rm c}_{\rm e}}\,
\left[
\mathcal{Q}(x_{\rm c}') 
- \Lambda_0\, \mathcal{Q}(x_{\rm c})
\right]
\id\mu \id\mu' 
\eta ^2\!\id \eta
\nonumber
\end{align}
where $\id^2 \sigma_2/\id \mu\id \mu'$ is defined by Eq.~\eqref{eq:sig_2} and we subtracted the term proportional to $\Lambda_0 =\id^2 \sigma^0_2/\id \mu\id \mu' / (\id^2 \sigma_2/\id \mu\id \mu')$ to cancel the temperature-independent contribution, $\frac{1}{30}\,\mathcal{Q}(x_{\rm c})$, caused by scattering of photons into the line-of-sight. Here $\id^2 \sigma^0_2/\id \mu\id \mu'$ denotes the cross section $\id^2 \sigma_2/\id \mu\id \mu'$ for $\beta=0$.

In Fig.~\ref{fig:kSZ_T_Q} we show the signals caused in second order of the cluster velocity.
Clearly, the correction already becomes very small, and if relative precision $\simeq10^{-4}-10^{-3}$ is not needed, these terms can certainly be neglected. For the signal related to the motion-induced quadrupole temperature corrections are again only visible at high frequencies, reaching $\simeq 10\%$ for $\The=0.05$.
The contribution to the SZ signal from the motion-induced monopole term vanishes in the limit $\The\rightarrow 0$, since Thomson scattering leaves the monopole unaltered.
As Fig.~\ref{fig:kSZ_T_Q} also indicates, the convergence of the expansions, Eq.~\eqref{eq:Y_kin} and \eqref{eq:Q_kin} is very slow, and good agreement is only reached for low electron temperature.
We confirmed numerically that the full 5-dimensional collision integral for the monopole and quadrupole correction is correctly represented by Eq.~\eqref{eq:BoltzEq_monopole_corr} and \eqref{eq:BoltzEq_quadrupole}.

In Fig.~\ref{fig:SZ_signals_comp} we illustrate the different SZ signals for $\The=0.05$, $\betac=0.01$ and some values of $\muc$.
For $\muc=0$ the kSZ signal vanishes and the dominant, but very small, corrections to the thSZ effect are second order in $\betac$. The motion-induced correction to the monopole is independent of $\muc$, as expected.
Also, for $\muc=-1$ the kSZ and qkSZ signals add while for $\muc=1$ they have opposite sign.
%

\begin{figure}
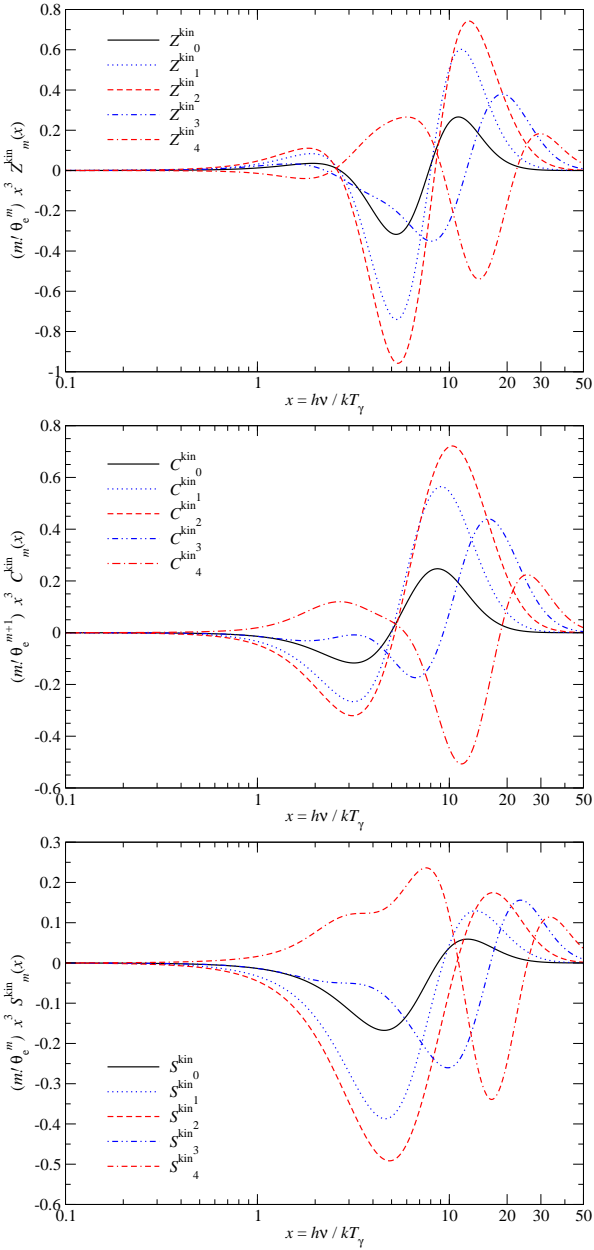

\centering
\includegraphics[width=0.93\columnwidth]{./eps/Zkin_functions.eps}
\\
\includegraphics[width=0.93\columnwidth]{./eps/Ckin_functions.eps}
\\ 
\includegraphics[width=0.93\columnwidth]{./eps/Skin_functions.eps}
\caption{First few functions $Z^{\rm kin}_k$, $C^{\rm kin}_k$, and $S^{\rm kin}_k$ for $\theta_{\rm e, 0}=0.03$. In {\sc SZpack} we computed up to $k=20$.
The functions $Z^{\rm kin}_k$ are needed to include temperature corrections related to the motion-induced monopole ($\propto \betacsq$), while $C^{\rm kin}_k$ and $S^{\rm kin}_k$ allow computing temperature corrections to the kSZ ($\propto \betac \muc$) and qkSZ effect ($\propto \betacsq P_2(\muc)$).
We rescaled by appropriate factors of $k$ and $\The$ to make them comparable in amplitude.}
\label{fig:basis_kin}
\end{figure}

\begin{figure}
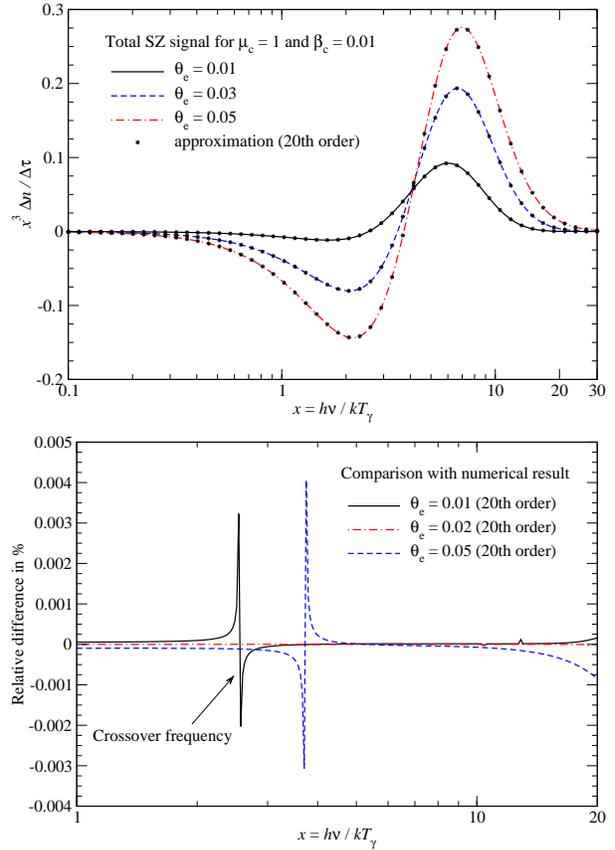

\centering
\includegraphics[width=0.94\columnwidth]{./eps/DI_SZ_all.eps}
\\[1mm]
\includegraphics[width=0.94\columnwidth]{./eps/DI_SZ_all.diff.eps}
\caption{Comparison of the expansion given by Eq.~\eqref{eq:thSZ_Zk_expression} and \eqref{eq:kSZ_ZCSk_expressionk} with the full numerical result for reference temperature $\theta_{\rm e, 0}=0.03$. The agreement clearly is very good. We considered $\betac=0.01$ and $\muc=1$ but found similar performance for other values of $\muc$.}
\label{fig:SZ_signals_comp_expansion}
\end{figure}

\subsubsection{Improved representation of kinematic SZ corrections}
\label{sec:k_basis_new}
The discussion of Sect.~\ref{sec:cluster_frame_dist} shows that the expansion in small frequency shifts for the kinematic corrections also suffers from the same limitation as the pure thSZ effect.
This problem again can be cured by defining a different set of basis functions to account for the various temperature terms:
\beal
\label{eq:Ilmk_b}
Z^{\rm kin}_k(x)&=   \frac{\mathcal{N}}{6 k!}\!\!
\int\!\! \frac{\id^2\sigma_0}{\id\mu \id \mu'}
\frac{\partial^k e^{-\Delta\gamma/\theta_{\rm e}^{\rm c}}}{\partial^k \The^{\rm c}}
\left[\mathcal{M}(x')-\mathcal{M}(x)\right]\id\mu \id\mu' \eta ^2\!\id \eta, 
\nonumber
\\[1mm]
C^{\rm kin}_k(x)&= 
-\frac{\mathcal{N}}{k!}\!\!
\int\!\! \frac{\id^2\sigma_1}{\id\mu \id \mu'}
\frac{\partial^k e^{-\Delta\gamma/\theta_{\rm e}^{\rm c}}}{\partial^k \The^{\rm c}}
\,\mathcal{G}(x')\id\mu \id\mu' \eta ^2\!\id \eta, 
\\[1mm]
\nonumber
S^{\rm kin}_k(x)&=   
\frac{\mathcal{N}}{3k!}\!\!
\int\!\! \frac{\id^2\sigma_2}{\id\mu \id \mu'}
\frac{\partial^k e^{-\Delta\gamma/\theta_{\rm e}^{\rm c}}}{\partial^k \The^{\rm c}}
\,\left[
\mathcal{Q}(x') 
- \Lambda_0\, \mathcal{Q}(x)
\right]
\id\mu \id\mu' \eta ^2\!\id \eta
\end{align}
%
Similar to the thSZ signal (Sect.~\ref{sec:T_basis_new}), we avoided taking the temperature derivative of the normalization factor, $\mathcal{N}(\The^{\rm c})$ by pulling it in front of the integral.
Using this basis, the kinematic corrections to the SZ signal in the small optical depth limit can be expressed as
\beal
\label{eq:kSZ_ZCSk_expressionk}
\Delta n^{\rm c}_{\rm kin}(\xgc,\vgh^{\rm c})
&\approx 
\betac \muc^{\rm c} \left(\Delta\tau^{\rm c}\mathcal{G}(\xgc)
+
\sum_{k=0}^{k_{\rm max}} \Delta z^{(k)} C^{\rm kin}_k(\xgc)
\right)
\\
\nonumber
&\!\!\!\!\!\!\!\!\!\!\!\!\!\!\!\!\!\!\!\!\!\!\!\!\!\!\!\!
+\betac^2\sum_{k=0}^{k_{\rm max}} \Delta z^{(k)} Z^{\rm kin}_k(\xgc)
+
\betacsq P_2(\muc^{\rm c})\left(
-\frac{3\Delta\tau^{\rm c}}{10}\mathcal{Q}(\xgc)
+\sum_{k=0}^{k_{\rm max}} \Delta z^{(k)} S^{\rm kin}_k(\xgc)
\right).
\end{align}
In Fig.~\ref{fig:basis_kin} we show the first few basis functions for $Z^{\rm kin}_k$, $C^{\rm kin}_k$, and $S^{\rm kin}_k$ for illustration. 
For precise representation of the SZ signal we numerically computed the basis up to $k_{\rm max}=20$ with relative error set to $\lesssim 10^{-6}$. 
In comparison to the normal asymptotic expansion the new set of basis functions again exhibits far fewer oscillations even for higher values of $k$. 
Equation~\eqref{eq:kSZ_ZCSk_expressionk} therefore allows very precise computation of the SZ signals for an observer at rest in the cluster frame.

\begin{figure*}
\centering
\includegraphics[width=1.4\columnwidth]{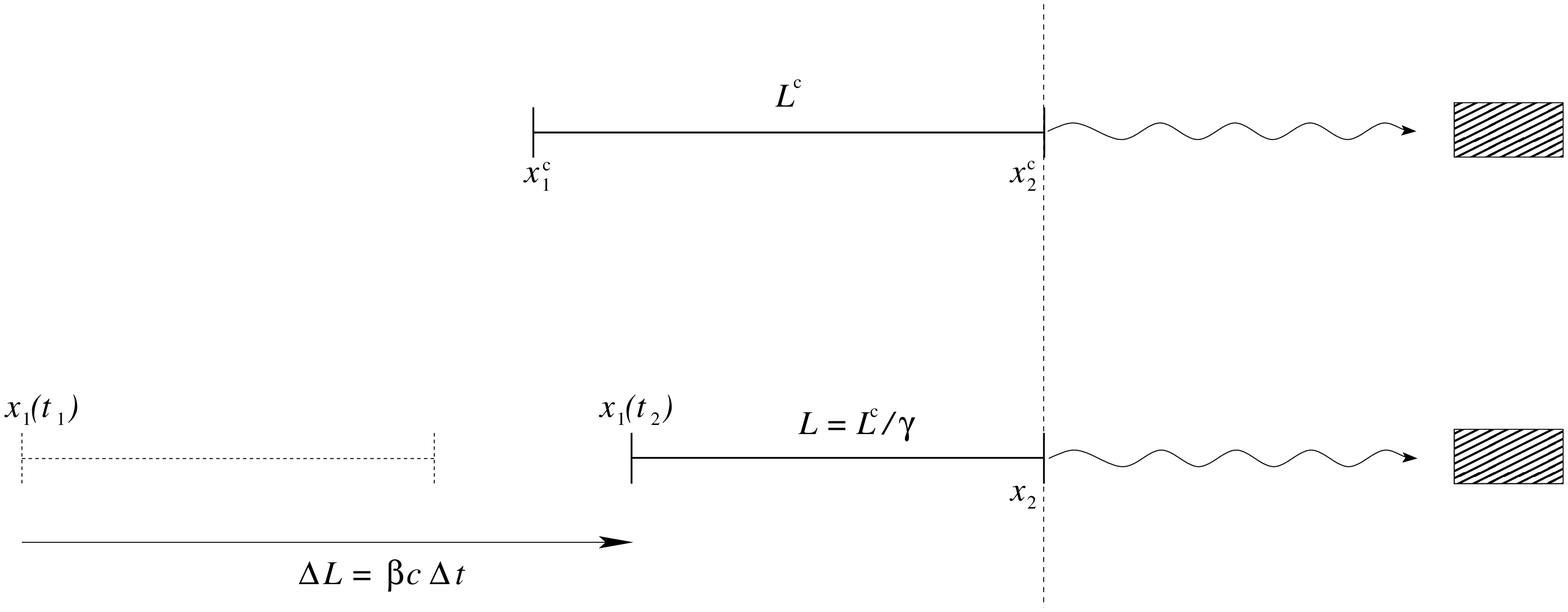}
\caption{
Illustration for the transformation of scattering optical depth. The received photons leave the scattering slab of medium at $x^{\rm c}_2=x_2$ and time $t^{\rm c}_2=t_2$. This is the space-time point at which we synchronized the wavefronts. 
The length of the slab measured inside the cluster frame is $L^{\rm c}=x^{\rm c}_2-x^{\rm c}_1=c\Delta t^{\rm c}$ (upper panel).
A photon that is scattered at $x^{\rm c}_1$ at time $t^{\rm c}_1=t^{\rm c}_2-\Delta t^{\rm c}$ into the line-of-sight is received by the observer at the same time as a photon that is scattered at $x^{\rm c}_2$ and time $t^{\rm c}_2$.
---
When the slab is moving with velocity $v=c\,\beta$ towards the observer (lower panel), the momentary length of the slab is reduced to $L=L^{\rm c}/\gamma$ by Lorentz-contraction.
A photon that is scattered at $x_1(t_1)=x_2(t_2)-(L+\Delta L)=c\Delta t$ is received by the observer at the same time as a photon scattered at $x_2(t_2)\equiv x^{\rm c}_2(t^{\rm c}_2)$ towards the distant observer.
During the interval $\Delta t$ the end of the slab moved by $\Delta L=\beta c \Delta t$ to $x_1(t_2=t^{\rm c}_2)=x_2-L$. We therefore find the effective time-interval $\Delta t=L/ c(1-\beta)=\Delta t^{\rm c}/\gamma(1-\beta)\equiv \gamma(1+\beta) \Delta t^{\rm c}$ over which the scatterings of photons that are received simultaneously by the resting observer take place.
For illustration we chose $\beta=0.6$, so that $L=0.8 L^{\rm c}$, $\Delta L=1.2 L^{\rm c}$ and $\Delta t=2\Delta t^{\rm c}$.
}
\label{fig:transform_Ill}
\end{figure*}

With Eq.~\eqref{eq:kSZ_ZCSk_expressionk} we can now compute the SZ signals in a very fast and economic way. 
In Fig.~\ref{fig:SZ_signals_comp_expansion} we illustrate the performance of this expansion. For all considered cases we find agreement at the level $\lesssim 0.001\%$ with the full numerical result in second order of $\betac$, at least away from the crossover frequency, where the total signal is small.
For temperatures $\The\lesssim 0.01$ one can use the asymptotic expansions given above to compute the SZ signals. A combination of both set of basis functions therefore allows calculation of the SZ signal with very high precision. 
We implemented these approximations in {\sc SZpack}, and found that even for $\The\simeq 0.06-0.07$ the new set of basis functions still represents the full numerical result with precision $\lesssim 0.1\%$.
We also checked the expressions for other values of $\muc$ and found excellent agreement with the full numerical result.

To emphasize the improvement in the performance, evaluation of the full 5-dimensional collision term on a standard laptop takes about 2 minutes for 40 frequency points with {\sc SZpack}, while calculation of the SZ signal with our improved set of basis functions is practically instantaneous. For computations of the SZ signals from simulated clusters this makes an important difference.

\subsubsection{Additional small corrections}
\label{sec:small_corrs}
When considering terms in second order of $\betac$, in reality one should also include the primordial CMB quadrupole, octupole and other large-scale anisotropies, which all have a spectrum $\xgc \partial_{\xgc} \nPl(\xgc)$. For clusters with $\betac\simeq \pot{3}{-3}$ these in principle have comparable amplitude ($\simeq 10^{-5}-10^{-4}$), and in different orders of the electron temperature the anisotropies they do couple to the Compton scattering cross section. 
With the 5-dimensional collision term we can estimate this effect using {\sc SZpack}.
We found that the correction always entered at the relative level of $\Delta n/n\lesssim 10^{-4}$, except for close to the crossover frequency, where the correction could reach $\simeq 0.1\%$ of the kSZ signal.
However, in absolute terms this correction is tiny.
Similarly, higher order motion-induced corrections remain small.  
If extremely high precision was required it would be straightforward to include these terms.
We also confirmed that the effect of stimulated scattering and electron recoil become important at a relative level $\Delta n/n \lesssim 10^{-6}$.

\subsection{Transforming the SZ signal into the CMB rest frame}
\label{sec:beta_problem}
The expressions for the SZ signal discussed so far are only valid for an observer at rest in the cluster frame.
Also, they describe the change of the photon occupation number caused by electron scattering assuming the optically thin limit, and multiplication by $2h\nu^3/c^2$ gives the distortion of the spectral intensity signal.
As mentioned above, in the cluster frame the derivation is rather straightforward and very clear in terms of the required scattering physics. 
However, the transformation to a general observer becomes slightly more complicated, especially when thinking about real measurements and the appearance of the cluster to the observer (see Sect.~\ref{sec:appear}).

Below we consider the transformation of the SZ intensity signal into an arbitrary observer frame. However, to understand the problem it is easiest to start with the special case of an observer at rest with respect to the CMB, where the CMB spectrum is given by a blackbody, $\nPl(x)$, with no anisotropies.
From the cosmological point of view this is also the most relevant frame, for instance when discussing about the statistics of the large-scale velocity field.
The first obvious transformation relates the frequencies of the observed photons: $\xgc=\gamma_{\rm c} x (1-\betac\muc)$, accounting for the Doppler effect. 
Transformation of $\vgh^{\rm c}$ gives $\muc^{\rm c}=(\muc-\betac)/(1-\betac \muc)$, describing the relativistic aberration effect.
Here $\muc=\vbetac\cdot \vgh$ is measured in the CMB frame, while the azimuthal angle under which the scattered photon is received does not change.
These expressions can be directly inserted into the formulae describing the SZ signals, e.g., Eq.~\eqref{eq:thSZ_Zk_expression} and \eqref{eq:kSZ_ZCSk_expressionk}, when using our improved set of basis functions.
However, there are two additional transformations that deserve a few comments. The first is related to the optical depth and the other to the electron and photon temperature.

\subsubsection{Transformation of the optical depth}
\label{sec:tau_transform}
Let us start by considering the optical depth along a given line-of-sight.
Because time along the photon's path transforms as\footnote{We denote time with $t$ and spatial coordinate with $\vek{r}$ in the CMB frame.} \citep[compare][]{Sazonov1998} 
$\id t^{\rm c}=\gamma^{-1}_{\rm c} (1+\betac\muc^{\rm c})^{-1} \id t 
\equiv \gamma_{\rm c} (1-\betac\muc) \id t$, for the differential of the optical depth we have $\id \tau^{\rm c}(t^{\rm c},\vek{r}^{\rm c})=\gamma_{\rm c} (1-\betac\muc)c \sigT \Ne^{\rm c}(t^{\rm c},\vek{r}^{\rm c}) \id t \equiv  (1-\betac\muc) \id \tau(t,\vek{r})$, where we used $\Ne^{\rm c}(t^{\rm c},\vek{r}^{\rm c})\equiv \Ne(t,\vek{r})/\gamma_{\rm c}$ \citep[also see][for more detailed explanation]{Chluba2012}.
The Lorentz factors cancel out after combining time-dilation and length-contraction.
The remaining factor of $(1-\betac\muc)$ accounts for the effect of retardation: while photons are propagating along a given line-of-sight the cluster is moving with $\betac\muc$ along that same line-of-sight. 
For a fixed interval $\Delta t\equiv \Delta t^{\rm c}$ this shortens the apparent distance that photons travel through the cluster medium by $(1-\betac\muc)$.

However, in terms of the SZ signals, the factor $(1-\betac\muc)$ should not be considered separately, since physically the relevant quantity is the total number of scattering events, $N_{\rm sc}(\vgh^{\rm c})$, along the photon path through the ICM, which is Lorentz-invariant and related to the integrated scattering optical depth \citep[see e.g.,][]{Rybicki1979, Abramowicz1991}. 
Different observers therefore always agree on the number of scatterings encountered by the photons along a given geodesic through the cluster. The apparent direction under which the photon is received along a particular geodesic and also the specific moments of the scattering events depends on the frame because of the aberration effect, time-dilation and also geometric effects, but $N_{\rm sc}(\vgh^{\rm c})$ is always identical.

To understand this point better, let us consider a simpler case in which a thin slab of scattering medium is moving with $\beta$ along the $x$-axis towards the observer (see Fig.~\ref{fig:transform_Ill} for illustration).
If the cluster and observer are at rest with respect to each other, then we have the optical depth $\Delta \tau^{\rm c}\equiv \int_{t^{\rm c}_1}^{t^{\rm c}_2} c \sigT \Ne^{\rm c} \id t^{\rm c} = \sigT \Ne^{\rm c} L^{\rm c} \propto N_{\rm sc}$. 
Here we assumed a constant number density, $\Ne^{\rm c}$, of electrons along the slab.
As indicated, $\Delta \tau^{\rm c}$ is directly proportional to the number of scatterings along the photon path, or equivalently, the number of electrons encountered by the photon along the given world line.

Using the transformation for $\id \tau$ given above, we also have $\Delta \tau^{\rm c}\equiv (1-\beta)\int_{t_1}^{t_2} c \sigT \Ne \id t = (1-\beta) c \sigT \Ne \Delta t$. 
As illustrated in Fig.~\ref{fig:transform_Ill}, the effective interval over which the scatterings take place increases to $\Delta t = \Delta t^{\rm c}/\gamma (1-\beta)=L^{\rm c}/c \gamma (1-\beta)$.
We therefore have $(1-\beta) c \sigT \Ne \Delta t = \sigT \Ne L^{\rm c}/\gamma = \Delta \tau^{\rm c}$, because $\Ne^{\rm c}=\Ne/\gamma$.
More generally, this implies that the integral $\Delta \tau^{\ast}(\vgh)=\int_{t_1}^{t_2} c \sigT \Ne \id t$ has the interpretation $\Delta \tau^{\ast}(\vgh)= \Delta \tau^{\rm c}(\vgh^{\rm c})/(1-\betac\muc)$ in terms of the cluster frame line-of-sight optical depth along the photon's world line defined by $\vgh^{\rm c}$.
This shows that $\Delta \tau^{\ast}(\vgh)$ is no longer just related to the number of scatterings (or the scattering optical depth in the classical sense) along the photon's path but includes a kinematic factor $(1-\betac\muc)^{-1}$ to account for the change of the electron flux in the CMB rest frame.
This implies that the variable $\Delta \tau^{\ast}(\vgh)$ describes a mixture of scattering effects and kinematic correction terms.

To compute the SZ signals in different observer frames we directly use $\Delta \tau^{\rm c}(\vgh^{\rm c})$, only transforming the apparent directions under which the photon are received.
This variable is very convenient because $\Delta \tau^{\rm c} \propto N_{\rm sc}$ without any extra dependence on the cluster's (or observer's) motion, such that a clean separation of scattering and kinematic effects is achieved.
As we explain in Sect.~\ref{sec:kSZ_Itoh}, in comparison with previous works \citep[e.g.,][]{Sazonov1998, Nozawa1998SZ, Nozawa2006, Shimon2004} this makes a difference for the interpretation of the kinematic correction terms at order $\mathcal{O}(\betac\The)$ and $\mathcal{O}(\betac^2)$.

\subsubsection{Transformations of electron and photon temperature}
The effects mentioned in this section only become relevant for future high, precision SZ observations, because they enter the SZ signal at order $\mathcal{O}(\betac \The)$ and $\mathcal{O}(\betacsq \The)$. Nevertheless, they touch on interesting aspects connected with the determination of the cluster frame electron temperature, $\Te^{\rm c}$, and the CMB temperature, $\Tg$, which are worth discussing briefly. 
In particular, when considering the problem of eliminating $\Te^{\rm c}$ from the expressions for the SZ signal, e.g., with X-ray observations, this section should be of interest.

It is evident that {\it only} in the rest frame of the moving cluster (or a small volume element with constant temperature and number density) it is possible to define $\Te^{\rm c}$ in a meaningful way \citep[see][for some related discussion]{Reiser1994, Juettnercov2010}.
In this sense the temperature of the gas becomes a parameter that, like the rest mass, describes the properties of a parcel of gas.
An observer, who is moving with respect to the cluster frame, can only {\it indirectly} infer the rest frame electron temperature, for example, by measuring the widths of thermally broadened resonance lines, the shape of the X-ray spectrum, or even by measuring the SZ signals discussed here.
In all these cases we observe some photons and by using the Lorentz-invariance of the photon occupation number, after making assumptions about the physical processes causing the signal, we can in principle understand how to infer the electron temperature defined in the rest frame of the cluster.
Consequently, the `measured' value for the electron temperature, $\Te'$, is directly linked to the method that was used to obtain it.

Thinking, for example, of a thermally broadened resonance, it is clear that to first order in $\betac$ only the position of the line changes, but its fractional width remains the same, 
so that one can directly infer $\Te' \approx \Te^{\rm c}$.
In second order of the motion this is no longer true, and the detailed transformation law for the photon distribution into the observer frame has to be considered.
When using the cluster's X-ray spectrum to determine the electron temperature, it is also important that to first order in $\betac$ the Doppler effect again changes the received photon frequency. This implies that the inferred value of the electron temperature (by fitting the shape of the X-ray spectrum) is $\Te' \simeq \Te^{\rm c}/(1-\betac \muc)$. However, in both the narrow line and X-ray spectrum case degeneracies with the redshift of the cluster exist.
This shows, that as long as kinematic corrections are insignificant (say in comparison to the measurement errors) we need not worry about the way $\Te^{\rm c}$ was inferred; however, small differences do arise at high precision, and dependencies on the method becomes an issue. Observationally, measurements of the electron temperature (e.g., from X-rays) at the sub-percent level are futuristic, so that this effect will not become significant any time soon.

For similar reasons, determinations of the monopole temperature of the CMB, $\Tg$, must be considered with care.
In the CMB rest frame, by averaging the spectrum over the whole sky, we directly find the value of $\Tg$, by comparing the average spectrum with a blackbody.
The same result is of course obtained by measuring the spectrum in any single direction (we neglect primordial CMB anisotropies).
However, for an observer who is moving with respect to the CMB, the `observed' value for the CMB temperature, $\Tg'$, in second order of the peculiar motion again depends on the observational strategy.
By subtracting the motion-induced dipole and quadrupole {\it temperature} anisotropy in the moving frame, the monopole temperature is $\Tg'\approx \Tg(1-\betac^2/6)$ \citep{Chluba2004}.
If on the other hand, we just integrate the observed {\it intensity} (including the motion-induced dipole and quadrupole {\it spectral} anisotropies) over the whole sky, we find an average energy density of $\rho_\gamma \approx a_{\rm R} \Tg^4(1+\frac{4}{3}\betac^2)$ because of the superposition of blackbodies with spatially varying temperatures \citep{Chluba2004, Chluba2012}.
Here $a_{\rm R}$ is the radiation constant. 
We therefore can infer an effective temperature of $\Tg'\approx \Tg(1+\betac^2/3)$.

These are just two simple (though not necessarily practical) examples of how the measurement 
procedure determines the relation between the observable, $\Tg'$, and the rest frame CMB temperature, $\Tg$.
Since in our expressions derived for the SZ signals both $\Te^{\rm c}$ and $\Tg$ appear, it is important to bear the above comments in mind, when interpreting measurements at high precision.
The values of $\Tg'$ and $\Te'$ obtained by some method in the moving frame affect the inference made for the SZ effect. 
For $\Tg$, this only enters the problem via terms connected with the thSZ effect, e.g., Eq.~\eqref{eq:thSZ_Zk_expression}, since all kinematic correction terms are already at least of first order in $\betac$.
However, for $\Te^{\rm c}$ kinematic terms can even give rise to corrections $\propto \betac \The$.
In other words, errors on the inferred values for $\Tg$ and $\Te^{\rm c}$ will become important when interpreting the SZ signals with high precision. 
For the current precision in the value of the CMB monopole, this limits the interpretation of SZ signals to no better than $\simeq 0.1\%$, but in the future the monopole of the CMB might be measured with much higher accuracy, for example using PIXIE \citep{Kogut2011PIXIE}.

\subsubsection{Final SZ signal in the CMB rest frame}
\label{sec:SZ_signal_CMB_frame}
With the comments above it is now straightforward to obtain the appropriate expressions for the SZ signal in the CMB rest frame.
Using Eq.~\eqref{eq:thSZ_Zk_expression} and Eq.~\eqref{eq:kSZ_ZCSk_expressionk}, for a given observing frequency, $x$, electron temperature, $\Te^{\rm c}$, cluster velocity, $\betac$ and direction, $\muc$, and line-of-sight optical depth, $\Delta \tau^{\rm c}$, we must evaluate the expressions at frequency $\xgc=\gamma_{\rm c} x (1-\betac\muc)$ and $\muc^{\rm c}=(\muc-\betac)/(1-\betac \muc)$. In this way we obtain a precise prediction for the associated SZ signal up to second order in $\betac$.
For simplicity one could insert $\muc^{\rm c}\approx \muc$ and $\xgc\approx x$ for all terms that are already of second order in $\betac$, but this is not required, as the `error' enters at $\mathcal{O}(\betac^3)$.
Similarly, for terms that are already first order in $\betac$ one can use $\muc^{\rm c}\approx \muc -\betac (1-\muc^2)$ and $\xgc\approx x (1-\betac \muc)$.

\begin{figure}
\centering
\includegraphics[width=0.94\columnwidth]{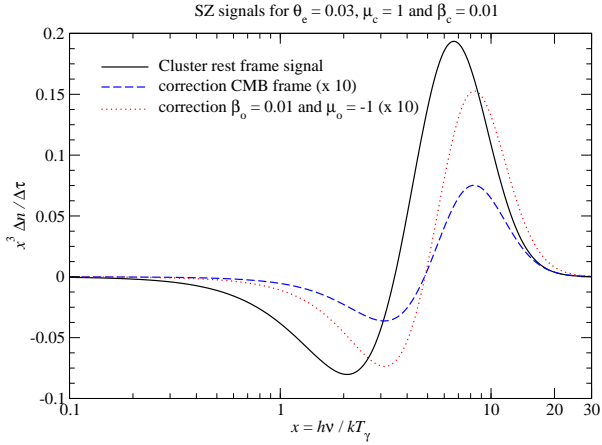}
\caption{Kinematic corrections from a change of the observer's frame. The solid/black line shows the SZ signal for a cluster with $\The = 0.03$, $\betac=0.01$ and $\muc=1$ in the cluster frame. The other two lines show the correction to this signal (times 10) when transforming into the CMB rest frame (dashed/blue line) and for an observer who is moving in the opposite direction of the cluster with $\betao=0.01$.}
\label{fig:SZ_signals_CMB_frame}
\end{figure}
In Fig.~\ref{fig:SZ_signals_CMB_frame} we illustrate how the transformation back into the CMB rest frame affects the observed signal.
The main effect is a shift of the photon spectrum towards higher frequencies.
For the considered case, the correction is comparable to $\simeq 5\%$ of the SZ signal in the cluster frame. In particular, close to the crossover frequency, $x_{\rm cr}\simeq 3.83$, the kinematic correction makes a large difference, shifting $x_{\rm cr}$ towards higher frequencies.

When looking at an extended object, there is a collection of sight lines through the object. Each of them has a different value of $\muc'$. 
Fixing the axis toward the object and assuming that it is moving as a whole in one direction with velocity vector defined by $\muc$ and $\phi_{\rm c}$ relative to this axis, $\muc'$ is given by $\muc'=\muc \mu + \cos(\phi_{\rm c}-\phi) \sqrt{(1-\muc^2)(1-\mu^2)}$. Here $\mu=\cos(\theta)$ and $\phi$ defines the line-of-sight relative to the reference axis. 
Typical clusters have angular size of only a few arcminutes ($\simeq \text{few} \times 10^{-4} \,{\rm rad}$), so we can approximate $\muc'\approx \muc(1-\frac{1}{2}\theta^2) + \cos(\phi_{\rm c}-\phi) \sqrt{1-\muc^2} \, \theta $. 
At a relative precision of $\simeq 10^{-4}-10^{-3}$ one can therefore neglect higher order terms and just use $\muc'\approx \muc$ for each line-of-sight through the cluster. This is equivalent to the {\it flat-sky approximation} and the optical depth in this case is just a function of position on the sky $\Delta \tau^{\rm c}(\theta, \phi)$ with roughly one common value of $\muc$ independent of $\phi_{\rm c}$.

\subsection{Transforming into the frame of a general observer}
\label{sec:trans_general_obs}
Following the discussion of the previous section, it is now trivial to give the transformation laws for a general observer.
If the observer is moving at a velocity $\vbetao$ with respect to the CMB, we can first transform the SZ signal from the cluster frame into the CMB frame and then subsequently relate the observed photon frequency $x_{\rm o}$ and photon direction $\vgh_{\rm o}$ to the corresponding quantities in the CMB frame.
To study the large-scale velocity fields we are interested in the values $\betac$ and $\muc$ inside the CMB frame, and it is therefore unnecessary to express these `parameters' in the observer frame.
Thus, we can simply use the transformations 
\beal
\label{eq:transformations_observ}
\xgc&=\gamma_{\rm c} x [1-\betac\muc] =\gamma_{\rm c} \gamma_{\rm o} x_{\rm o} [1-\betac\muc] [1+\betao\muo],
\nonumber
\\
\muc^{\rm c}&=\frac{\muc-\betac}{1-\betac \muc},
\qquad\qquad \mu=\frac{\muo+\betao}{1+\betao \muo},
\end{align}
with the Lorentz-factors $\gamma_{\rm c}=1/\sqrt{1-\betac^2}$ and $\gamma_{\rm o}=1/\sqrt{1-\betao^2}$, and direction cosines $\muc=\vbetach \cdot \vgh$, $\mu=\vbetaoh \cdot \vgh$, and $\muo=\vbetaoh \cdot \vgh_{\rm o}$.
The photon frequency, $x_{\rm o}$, and the value of $\muo$ are directly measured in the observer's frame, but the dependences of the SZ signal on the cluster's and observer's peculiar motions and the relevant direction cosines are still expressed in the rest frame of the CMB.
As mentioned above, this is a reasonable choice, since in this way the interpretation of the measured SZ signal (in the observer's frame) directly reveals $\betac$ and $\muc$ in the CMB rest frame, the quantities we are after from the cosmological point of view.

The value of $\betao$ can be directly determined by looking at the CMB temperature dipole inside the observer's frame. 
For our motion with respect to the CMB rest frame we have $\betao=\pot{1.241}{-3}(1\pm 0.2\%)$ towards $(l, b)=(264.14^\circ\pm 0.15^\circ, 48.26^\circ\pm 0.15^\circ)$ in galactic coordinates \citep{Fixsen1996, Fixsen2002}.
The value of $\muo$ is then also fixed, by comparing the position of the cluster on the sky with the dipole axis.
Since $\betao$ is small the additional modification caused by our motion relative to the CMB rest frame also remains small. This effect was first investigated by \citet{Chluba2005b}. Their expressions for the SZ intensity change were confirmed by \citep{Nozawa2005}. It was also pointed out that the motion of the Solar System should lead to a dipolar anisotropy in the cluster number counts \citep{Chluba2005b}. 
Furthermore, the small errorbars on $\betao$ and the direction of the dipole imply that the effect of our motion with respect to the CMB can be taken out with $\simeq 0.1\%$ precision.

With Eq.~\eqref{eq:transformations_observ}, we can verify that $\xgc=x_{\rm 0}$ for $\vbetac \equiv \vbetao$. Here it is important to note that in the observer's frame we measure $\muo\equiv \muc^{\rm c}$, and hence $(1-\betac\muc)(1+\betao\muo)=(1-\betac\muc)(1+\betac\muc^{\rm c})\equiv \gamma_{\rm c}^{-2}$.
For an observer at rest in the CMB frame we also find $x_{\rm o}=x$ and $\mu_{\rm o}=\mu$, as expected.
Note that by using the expressions~\eqref{eq:transformations_observ} we can avoid additional approximations related to the transformation into the observer's frame. The precision of the predicted SZ signal is therefore only limited by how well we are able to represent the distortion in the cluster frame.

In Fig.~\ref{fig:SZ_signals_CMB_frame} we illustrate how the SZ signal is affected by the transformation to an observer moving with a rather large velocity in the opposite direction of the cluster.
For comparison, our motion relative to the CMB rest frame causes a maximal correction that is about one order of magnitude smaller.
Notice that for the computation of the SZ signal we used our expansions, Eq.~\eqref{eq:thSZ_Zk_expression} and Eq.~\eqref{eq:kSZ_ZCSk_expressionk}, and only needed to fix $\Delta \tau $, $\Te$, $\Tg$, $\betac$, $\muc$, $\betao$, and $\muo$.
Here, we assumed that the cluster is isothermal along the line-of-sight and we omitted bulk flows of the ICM. 
From CMB observations we can directly infer $\Tg$, $\betao$, and $\muo$. The remaining parameters describing the SZ signal are $\Delta \tau$, $\Te$, $\betac$, and $\muc$; these can be directly constrained by multi-frequency SZ observations, where $\Delta \tau$ and $\Te$ now denote the cluster frame scattering optical depth and electron temperature, respectively.
Generalization to the non-isothermal case and inclusion of bulk flows of the ICM is straightforward and will be discussed in our future work.

\subsection{Comparison with previous results (CMB rest frame)}
\label{sec:kSZ_Itoh}
We now compare our results for the kinematic corrections directly with the expressions given by different groups.
For this we consider the SZ signal as measured in the CMB rest frame.
Following the discussion of Sect.~\ref{sec:beta_problem}, and using the analytic expressions for the asymptotic expansion given above, for $\betac\ll 1$ we find
\beal
\label{eq:SZ_signal_CMB}
\frac{\Delta n(x,\vgh)}{\Delta\tau(\vgh)}
&\approx 
 \The \! \sum_{k=0}^{k_{\rm max}} \The^{k} \left(Y_k+\betacsq 
 \left[ Y^{\rm kin}_k+\frac{1}{6} \mathcal{D}_{x} Y_k
 \right]
 \right) 
 -\frac{1}{3}\betacsq \mathcal{D}^\ast_{x} \mathcal{K}(x)
\nonumber 
\\
&\!\!\!\!\!\!\!\!\!\!\!\!
+ \betac \muc \left(\mathcal{G} +  \The\sum_{k=0}^{k_{\rm max}} \The^{k} 
\left[D^{\rm kin}_k - x \, \partial_x Y_k \right] \right)
\\
\nonumber 
&\!\!\!\!\!\!\!\!
+\betacsq P_2(\muc)\left( -\frac{3}{10}\mathcal{Q} 
+\frac{2}{3}\mathcal{D}^{\ast\ast}_{x}  \mathcal{K} 
+\The\sum_{k=0}^{k_{\rm max}} \The^{k} \left[ Q^{\rm kin}_k  
+ \frac{1}{3} x^2 \partial^2_x Y_k \right] \right).
\end{align}
We emphasize again that $\Delta \tau$ and $\Te$ denote the {\it cluster frame} scattering optical depth and electron temperature, respectively.
All frequency-dependent functions are evaluated at frequency $x$ and we used $\mu^2 = \frac{2}{3} P_2(\mu)+\frac{1}{3}$.
We also defined the differential operators $\mathcal{D}_{x}=3 x\, \partial_x + x^2 \partial^2_x$, $\mathcal{D}^\ast_{x}=2+ x\, \partial_x$, and $\mathcal{D}^{\ast\ast}_{x}=1-x\, \partial_x$, and the spectral function 
$\mathcal{K}=\left(\mathcal{G}  + \The\sum_{k=0}^{k_{\rm max}} \The^{k} D^{\rm kin}_k \right)$.
As before, we assumed that the cluster is isothermal and has no internal bulk flows.

With Eq.~\eqref{eq:SZ_signal_CMB} we can easily read off terms of different orders in the electron temperature and $\betac$.
As mentioned earlier, our expression for the purely thSZ effect agree with those given earlier in the literature \citep{Challinor1998, Sazonov1998, Itoh98, Shimon2004}.
Also the kSZ term, $\propto \betac\muc \mathcal{G}$, is the same.
For the qkSZ effect we however find 
\beal
\label{eq:SZ_signal_CMB_P2_0}
\frac{\Delta n^{\rm qkSZ}(x,\vgh)}{\Delta\tau(\vgh)}
&\approx  \frac{11}{30}\,\betacsq P_2(\muc) \mathcal{Q}(x)
\end{align}
in the CMB frame, where we used $\mathcal{D}^{\ast\ast}_{x}  \mathcal{G} \equiv 
\mathcal{Q}(x) = \mathcal{G}(x) \,\tilde{X}(x)$.
This result differs by $-\frac{2}{3}\betacsq P_2(\muc) \mathcal{G}(x)$ from the corresponding terms given in\footnote{\citet{Challinor1999} only gave the kinetic equation to describe the scattering of CMB photons by moving electrons, but did not integrate it.} \citet{Sazonov1998}, \citet{Nozawa1998SZ}, \citet{Shimon2004}, and \citet{Nozawa2006}.

The difference arises because other conventions for the optical depth variable were used in these earlier works.
This aspect was not explicitly addressed previously, but with the explanations given in Sect.~\ref{sec:tau_transform} it is straightforward to show that \citet{Sazonov1998} and \citet{Shimon2004} expressed the SZ signal in terms of $\Delta \tau^{\ast} = \Delta \tau / (1-\betac \muc)$, while \citet{Nozawa1998SZ} and \citet{Nozawa2006} used $\Delta \tau^{\ast\ast} = \Delta \tau^{\ast}/\gamma_{\rm c}$.
These redefinitions explain the above difference and consequently imply alternative interpretations of the kinematic corrections to the SZ signal in each case.
From the physical point of view the variables $\Delta \tau^{\ast}$ and $\Delta \tau^{\ast\ast}$ both describe a mixture of scattering and kinematic terms, while with our choice of variables these effects are cleanly separated.

Similarly, for the kinematic term $\propto \betacsq$ (with no additional dependence on $\muc$) we find
\beal
\label{eq:SZ_signal_CMB_M_0}
\frac{\Delta n^{\rm mkSZ}(x,\vgh)}{\Delta\tau(\vgh)}
&\approx  \frac{1}{3}\,\betacsq \left[\mathcal{Q}(x)-3\mathcal{G}(x) \right]
\equiv \frac{1}{3}\,\betacsq \left[ Y_0(x) + \mathcal{G}(x) \right]
\end{align}
using $\mathcal{D}^{\ast}_{x}  \mathcal{G} \equiv - \mathcal{Q}(x) + 3G(x)=-\mathcal{M}(x)$.
This differs by $-\frac{\betacsq}{3} G(x)$ from the results of \citet{Sazonov1998}, \citet{Nozawa1998SZ}, \citet{Shimon2004}, and \citet{Nozawa2006}, for the same reason as above.

\subsubsection{Temperature and velocity cross-term in first order of $\betac$}
From Eq.~\eqref{eq:SZ_signal_CMB} in first order of the electron temperature we obtain
\beal
\label{eq:SZ_signal_CMB_bT_X}
\frac{\Delta n^{\rm X}(x,\vgh)}{\Delta\tau(\vgh)}
\nonumber
&\approx \The \betac \muc \left[D^{\rm kin}_0 - x \, \partial_x Y_0 \right]
\\
&=\The \betac \muc \left[-\frac{138}{5}\mathcal{G}-\frac{42}{5}Y_0
+\frac{7}{5}x\mathcal{G}(6\mathcal{G}+x)\right].
\end{align}
Again this expression differs from the one of \citet{Sazonov1998} and \citet{Nozawa1998SZ, Nozawa2006} by the amount expected from the redefinitions of the optical depth variables explained above, i.e., $-\The \betac \muc Y_0$.
The result of \citet{Shimon2004} for this term differs from ours by additional $-\The \betac \muc \mathcal{G}$. We attribute this to a slight inconsistency related to the use of the Lorentz-invariance of the electron phase space distribution in \citet{Shimon2004}, but did not investigate this question any further.
This difference was also discussed in \citet{Nozawa2006}.

To conclude, at first order in $\betac$ we find full agreement of our results for the asymptotic expansion of the Boltzmann collision term with those given in previous works \citep{Sazonov1998, Nozawa1998SZ, Nozawa2006}; however, only after including the extra factor of $(1-\betac \muc)$ caused by the retardation effect in the definitions of the scattering optical depth. This factor is absorbed by the line-of-sight integral and does not appear separately, as explained in Sect.~\ref{sec:tau_transform}.
When interpreting the SZ signals of individual clusters this difference can become important at the level of a few percent.

\subsubsection{Temperature and velocity cross-term in second order of $\betac$}
For similar reasons, the temperature correction $\propto \betacsq \The P_2(\muc)$ given by \citet{Nozawa1998SZ, Nozawa2006} and \citet{Shimon2004}, and all other second order kinematic terms also differ from our expressions by some small amount. 
For example, our first order temperature correction $\propto \betacsq\The P_2(\muc)$ reads
\beal
\label{eq:SZ_signal_CMB_qkSZ_The}
\frac{\Delta n^{\rm qkSZ, (1)}(x,\vgh)}{\Delta\tau(\vgh)}
&\approx  \betacsq \The P_2(\muc) 
\left[ 
Q_0^{\rm kin}+\frac{2}{3}(1-x\,\partial_x) D^{\rm kin}_0 + \frac{2}{3} x^2 \partial^2_x Y_0
\right]
\nonumber
\\
&\!\!\!\!\!\!\!\!\!\!\!\!\!\!\!\!\!\!\!\!\!\!\!\!\!\!
=\betacsq \The P_2(\muc) 
\left[ \frac{128}{5}\mathcal{G}+\frac{32}{5}Y_0
+6\xg^3 \partial_x^3 \nPl
+\frac{19}{30}\xg^4 \partial_x^4 \nPl
\right],
\end{align}
where we used Eq.~\eqref{eq:DQ_kin_examples} to simplify the expression.
The additional term related to the retardation factor $(1-\betac \muc)$ from the optical depth gives rise to $-(2/3)\betacsq \The P_2(\muc) ( D^{\rm kin}_0 - x\,\partial_x Y_0)$, yielding
\beal
\frac{\Delta n^{\rm qkSZ, (1)}(x,\vgh)}{\Delta\tau^{\ast}(\vgh)}
&\approx\betacsq \The P_2(\muc) 
\left[ 44\mathcal{G}+12 Y_0
+6\xg^3 \partial_x^3 \nPl
+\frac{19}{30}\xg^4 \partial_x^4 \nPl
\right].
\nonumber
\end{align}
With Eq.~\eqref{eq:relations_nPl_Itoh} one can show the equivalence to the expression of \citet{Nozawa1998SZ}, noting that $\Delta\tau^\ast=\Delta\tau^{\ast\ast} + \mathcal{O}(\betacsq)$.
However, in the present work the factor $(1-\betac \muc)$ is not considered separately, but as part of the optical depth integral.
Otherwise, the interpretation of optical depth is different and does not only reflect the effect of scattering, as explained above.

\citet{Nozawa1998SZ} did not explicitly carry out the transformation of the electron number density into the CMB rest frame. This introduces another factor of $\gamma^{-1}_{\rm c}\approx 1-\frac{1}{2}\betacsq$, which affects the expressions
for the terms $\propto \betacsq \The^k$ (without extra $\muc$-dependence).
For example, we find
\beal
\label{eq:SZ_signal_CMB_mqkSZ_The}
\frac{\Delta n^{\rm mkSZ, (1)}(x,\vgh)}{\Delta\tau(\vgh)}
&\approx  \betacsq \The
\left[ 
2Y_0^{\rm kin}-\frac{2}{3}(2+x\,\partial_x) D^{\rm kin}_0 
\right]
\\
\nonumber
&\!\!\!\!\!\!\!\!\!
=\betacsq \The
\left[ 
\frac{234}{5}\mathcal{G}+\frac{66}{5}Y_0
+\frac{77}{15}\xg^3 \partial_x^3 \nPl
+\frac{7}{15}\xg^4 \partial_x^4 \nPl
\right],
\end{align}
where we made use of Eq.~\eqref{eq:Y_kin_examples}, and $\mathcal{D}_x Y_k\equiv 6 Y_k^{\rm kin}$.
Transforming to $\Delta\tau^\ast$ gives an extra term, $-\frac{1}{3}\betacsq \The [D^{\rm kin}_0-x\,\partial_x Y_0]$, so that
\beal
\frac{\Delta n^{\rm mkSZ, (1)}(x,\vgh)}{\Delta\tau^{\ast}(\vgh)}
&\approx
\nonumber \betacsq \The
\left[\frac{1}{3}  Y_0 + \frac{2}{3} Y_1
\right].
\end{align}
This is precisely $\frac{1}{2}\betacsq \The Y_0$ smaller than the expression of \citet{Nozawa1998SZ}, which, as mentioned above, is because $\Ne^{\rm c}$ was used to define the optical depth integral instead of $\Ne=\Ne^{\rm c}/\gamma_{\rm c}$.
However, this modification is small and should not affect the final result and interpretation of the SZ signal significantly.
We also note that the difference in the definition of $\Ne$ does not affect the lowest order $\betacsq$ term, Eq.~\eqref{eq:SZ_signal_CMB_mqkSZ_The}, since Thomson scattering leaves the CMB monopole spectrum unaltered.

\subsubsection{Energy-transfer and conservation of photon number}
\label{sec:test_results}
From the final solution for the cluster SZ signal one can compute the change in the number density of photons and also the total Compton energy transfer caused by the moving cluster.
This can directly serve as a consistency check \citep{Sazonov1998, Itoh98}, because the total number of photons should not change and also the energy transfer rate is well-known in this case.
However, when considering different lines-of-sight through a extended object and the average properties of the received photon intensity, some subtle effects arise which we explain here.

Before carrying out the line-of-sight integral, from the photon Boltzmann equation we have
\beal
\label{eq:SZ_Collision_term}
\frac{\id n(x,\vgh)}{\id t}
&\approx 
(1-\betac \muc) c \sigT \Ne \frac{\Delta n(x,\vgh)}{\Delta\tau(\vgh)},
\end{align}
where $\Delta n(x,\vgh)/\Delta\tau(\vgh)$ is given by Eq.~\eqref{eq:SZ_signal_CMB}.  
This expression is consistent with \citet{Sazonov1998} for all orders in $\betac$ and $\The$ presented there.
Up to second order in $\betac$, it is also equivalent to the expression obtained from the Lorentz-boosted Boltzmann collision term, which was used by \citet{Nozawa1998SZ} and more recently in the cosmological context by \citet{Chluba2012}. However, as explained above the collision term presented by \citet{Nozawa1998SZ} has to be multiplied by $1/\gamma_{\rm c}$ in addition, to transform $\Ne^{\rm c}$ to $\Ne$.

For a fixed time, from Eq.~\eqref{eq:SZ_Collision_term} we can carry out the angle averages and compute how much the photon field is affected by scatterings at a given location. 
Since $\frac{1}{2}\int P_i(\mu) \id \mu = 0$ for $i>0$, and because $\frac{1}{2} \int \mu P_1(\mu) \id \mu = \frac{1}{3}$ and $\frac{1}{2} \int \mu P_2(\mu) \id \mu = 0$, we find
\beal
\label{eq:SZ_Collision_term_av}
\left<\frac{\id n(x,\vgh)}{c \sigT \Ne\id t}\right>
&\approx 
 \The \! \sum_{k=0}^{k_{\rm max}} \The^{k} 
 \left(Y_k
 +2\,\betacsq   Y^{\rm kin}_k
 \right) 
 -\frac{1}{3}\betacsq \mathcal{D}^\ast_{x} \mathcal{K}(x)
\nonumber 
\\
&\qquad
-\frac{1}{3} \betacsq \left(\mathcal{G} +  \The\sum_{k=0}^{k_{\rm max}} \The^{k} 
\left[D^{\rm kin}_k - x \, \partial_x Y_k \right] \right),
\end{align}
Considering, for example, the case $\The=0$
we have
\beal
\label{eq:SZ_Collision_term_av_energy}
 \left<\frac{\id n(x,\vgh)}{c \sigT \Ne\id t}\right>_{\The=0}
&\approx 
 \frac{1}{3}\betacsq \mathcal{Q}(x)
 -\frac{4}{3} \betacsq \mathcal{G} \equiv  \frac{1}{3}\betacsq Y_0(x).
\end{align}
This expression implies $\left<\frac{\id N_\gamma}{\id t}\right>_{\The=0}=0$ for the change of the local photon number density, confirming that the scattering event conserves photon number. 
We also find $\left<\frac{\id \rho_\gamma(\vgh)}{\id t}\right>_{\The=0}=\frac{4}{3} \betacsq \rho_\gamma c \sigT \Ne$ for the change of the energy density, which is in agreement with the classical result for the Compton energy transfer \citep[e.g., see][]{Blumenthal1970, Rybicki1979}.
We confirmed that higher order terms also yield consistent results with respect to photon number conservation and energy transfer.

However, an important difference arises if we {\it first} carry out the line-of-sight integral and then average over different photon directions. 
Let us again consider the terms for $\The=0$. From Eq.~\eqref{eq:SZ_signal_CMB} and \eqref{eq:SZ_signal_CMB_M_0} we have
\beal
\label{eq:SZ_time_av_first_ang}
 \left<\frac{\Delta n(x,\vgh)}{\Delta \tau(\vgh)}\right>_{\The=0}
&\approx 
\frac{1}{3}\betacsq Y_0(x) + \frac{1}{3}\betacsq \mathcal{G},
\end{align}
which means
\beal
\label{eq:SZ_time_av_first_ang_num_energ}
 \left<\frac{\Delta N_\gamma}{\Delta \tau}\right>_{\The=0}
&\approx \betacsq N_\gamma
\qquad{\text{and}}
& \left<\frac{\Delta \rho_\gamma}{\Delta \tau}\right>_{\The=0}
\approx \frac{8}{3}\betacsq \rho_\gamma.
\end{align}
%
Not only does this apparently violate the photon number conservation, but also the transferred power is two times larger than that given by the classical formula. What is happening here?
The explanation is related to the difference between {\it emitted} and {\it received} power from a moving source \citep{Rybicki1979}.
For the classical expression given above, the power emitted at a single point is integrated over all directions for any given instance of time.
Imagining several sources embedded along a slab of medium, the average emitted power from all these sources is proportional to the number of sources inside the slab. However, the total power emitted by these sources was computed for each individual source separately and then averaged over the different sources. 
This is equivalent to placing all the sources into a single point and then averaging the emitted photon field using time-intervals measured in the source frame.
In contrast, the averages Eq.~\eqref{eq:SZ_time_av_first_ang_num_energ} are carried out such that the time-intervals are measured by the observer, with the retardation effects consistently included.
For SZ observations by some distant observers placed on a sphere around the cluster, Eq.~\eqref{eq:SZ_time_av_first_ang_num_energ} is therefore the relevant expression.

\subsection{Appearance of the SZ cluster and additional kinematic effects on the measured quantities}
\label{sec:appear}
Thus far we have only considered the transformation of the photon occupation number and how the spectral intensity along a given line-of-sight is affected by kinematic corrections.
However, additional effects arise in practice.
For example, due to the aberration effect different rays of photons change their separation: in the direction of the motion they become closer, while in the opposite direction they diverge from each other.
This means that when observing the flux of photons in a given direction and some fixed solid angle in different frames, the signal changes accordingly.
Consequently, the apparent angular size of the cluster changes by a small amount \citep{Chluba2005b}, depending on the direction of the motion of the observer relative to the cluster.
Furthermore, the geometry of the cluster affects the apparent shape on the sky, and deviations from non-isothermality and the radial distributions of electrons modify the SZ signals.
If we were to compare the results obtained by different observers or in different directions on the sky, this effect must be taken into account.

Also, in observations the photon flux is integrated over some frequency filter and with a finite sampling rate. Fixing these in different frames implies kinematic corrections to the `observed' SZ signals for different observers.
Similarly, it matters if photon flux or intensity are consider.
We will discuss these aspects of the problem in more detail in another work.

\subsection{Astrophysical contaminations}
\label{sec:contaminations}
There is a long list of additional effects that affect the SZ signal or compromise its interpretation at some significant level. For example,
we neglected the effect of multiple scattering \citep[see e.g.][]{Rephaeli1995ARAA, Molnar1999, Dolgov2001, Shimon2004, Nozawa2009}, which should be most noticeable close to the crossover frequency.
It is also known that the ICM should host populations of non-thermal electrons \citep[e.g.,][]{Ensslin2000, Blasi2000, Colafrancesco2003}. These give rise to non-thermal SZ distortions, which we omitted here, but do affect the SZ signal, especially at high frequencies. 
Also, the local radiation field inside the cluster is not necessarily described by a pure blackbody but could be contaminated, e.g., by synchrotron radiation. In that case one must also consider the effect of up-scattering for these parts of the photon distribution, another problem we leave for some future investigation.
Moreover, galaxy clusters can act as a moving lens, deflecting photon from the CMB background and causing a characteristic temperature pattern which can reach a level comparable to $\simeq 10\%$ of the kSZ effect \citep[e.g.,][]{Birkinshaw1983, Molnar2003}.
Finally, we did not account for the effects of the inhomogeneous ICM structures and a suite of astrophysical processes, such as turbulent gas motions \citep[e.g.,][]{Dolag2005, Lau2009, Vazza2009, Nelson2011} or magnetic fields \citep[e.g.,][]{Koch2003, Gopal2010JCAP}, on the SZ signal.
For accurate predictions and interpretation of the SZ signals {\it all} these processes must be included.

\section{Summary}
\label{sec:conc}
We considered the SZ signals caused by the scattering of CMB photons by hot electron residing inside clusters of galaxies, obtaining improved expressions that allow fast and precise computation of these signals.
We compute the effect in the cluster frame and then derive the signal for the general observer by Lorentz-transformation.
This procedure is physically well-defined and allows us to understand the origin of the different kinematic correction terms in a simple way. 
The method provides a quasi-exact representation of the Compton collision term, with frequency-dependent basis functions that are informed by the underlying physics of the scattering process.
The precision of the calculation is limited by the accuracy achieved inside the cluster frame, which we push well beyond the precision of current SZ observations and uncertainties associated with cluster physics (see Sect.~\ref{sec:contaminations}).

As explained here, the kinematic corrections to the SZ intensity signal obtained in previous analyses differ from ours by a small (comparable to a few percent for a typical cluster) amount (Sect.~\ref{sec:kSZ_Itoh}). This is because here we explicitly express the SZ signal in terms of the cluster frame optical depth along the photon's path. This choice of variable provides a clean separation of kinematic effect from those related to scattering events. The difference might be important for future precision measurements of the clusters bulk velocity and when confirming the redshift-scaling of the CMB temperature using the SZ effect.
In Sect.~\ref{sec:appear} we also briefly mention additional kinematic effects on the appearance of the SZ cluster and related instrumental issues; however, a more detailed analysis is beyond the scope of this paper.

One of the main product of this work is {\sc SZpack}, a numerical library which allows fast and precise ($\lesssim 0.001\%$ at frequencies $h\nu \lesssim 20 k\Tg$) computation of the SZ signals up to high electron temperature ($k\Te \simeq 25\,{\rm keV}$) and large peculiar velocity ($v/c \simeq 0.01$).
{\sc SZpack} should be useful when analyzing the SZ signals of high resolution, multi-frequency observation carried out, for example, with {ALMA} \citep{ALMA2008} and {CCAT} \citep{CCAT2006}.
Furthermore, the package should enable fast and accurate calculations of the predicted SZ signals from high-resolution clusters simulations.
Extensions to account for the effect of non-thermal electrons using the method described here are planned. 
Similarly, our approach should be applicable when polarization effects \citep[see, e.g.,][]{Itoh2000Pol, Challinor2000} are considered.
The impact of other processes mentioned in Sect.~\ref{sec:contaminations} will be studied more carefully in our future work.

\section*{Acknowledgments}
The authors cordially thank Anthony Challinor for careful refereeing of the paper and detailed comments that greatly helped improving the manuscript.
They also thank Prof. Itoh and Prof. Nozawa for helpful correspondence and suggestions.
JC in particular thanks Eric Switzer for his comments and insightful remarks on the paper.
He is also very grateful for useful discussions with Dimitrios Giannios, Takeshi Kobayashi, Richard Shaw and Geoff Vasil, which helped clarifying issues related to the transformation of the scattering optical depth along the world line of photons.
Furthermore, the authors acknowledge the use of the GPC supercomputer at the SciNet HPC Consortium. 
SciNet is funded by: the Canada Foundation for Innovation under the auspices of Compute Canada; 
the Government of Ontario; Ontario Research Fund - Research Excellence; and the University of Toronto.
DN and KN acknowledge the support from the NSF grant AST-1009811, the NASA  ATP grant NNX11AE07G, the NASA Chandra Theory grant GO213004B, and by Yale University.

\begin{appendix}

\section{Computing derivatives of the Planck spectrum}
\label{app:nPl_derives}
The derivatives of a Planckian distribution can be represented in different ways. To compute the temperature corrections to the thSZ effect up to $n^{\rm th}$ order of $\The$ we need derivatives up to order\footnote{To analytically include all kinematic corrections in $\mathcal{O}(\betac^2)$ two additional derivatives are needed.} $k=2n+2$. 
Here we give explicit expressions for $k\leq 8$; however, we also show how to compute these derivatives up to any order using {\it Eulerian numbers}.
This procedure is useful, since the expressions are rather lengthy while the relations using Eulerian numbers are basically analytic, involving summation of previous coefficients.
Also, the recursions given here greatly increase the numerical stability for the evaluation of higher order derivatives.

\subsection{Explicit expressions up to the $8^{\rm th}$ derivative of $\nPl$}
\label{app:nPl_derives_Itoh_GX}
%
In \citet{Itoh98} and \citet{Sazonov1998} the derivatives $\xg^k \partial^k_{\xg} \nPl$ were expressed using the variables
\beal
\tilde{X}(x)&=x \coth\left(\frac{x}{2}\right),
\qquad 
\tilde{S}(x)=x/\sinh\left(\frac{x}{2}\right).
\nonumber
\end{align}
To find the corresponding expressions for different derivatives of the Planck function, $\nPl(x)=1/(e^x-1)$, the relations 
\beal
x^{k+1} \partial^{k+1}_{x}&=\left(x \partial_{x}-k\right)x^{k} \partial^k_{x}
\nonumber\\
x \, \partial_{x}\tilde{X} &=  \tilde{X} -\frac{1}{2} \tilde{S}^2
\nonumber\\
x \, \partial_{x}\tilde{S} &=  \tilde{S}-\frac{1}{2} \tilde{X}\tilde{S}
\nonumber
\end{align}
are very useful. 
For example, in terms of $\tilde{X}$ and $\tilde{S}$ one can write
\beal
\label{eq:relations_nPl_Itoh}
\xg\partial_{\xg} \nPl &=-\mathcal{G},
&
\xg^2\partial^2_{\xg} \nPl &=\mathcal{G} \,\tilde{X},
\nonumber\\
\xg^3\partial^3_{\xg} \nPl &=-\mathcal{G}\left[\tilde{X}^2 +\frac{1}{2} \tilde{S}^2 \right],
&
\xg^4\partial^4_{\xg} \nPl &=\mathcal{G}\,\tilde{X}\,\left[\tilde{X}^2 +2\tilde{S}^2 \right],
\end{align}
where in addition we introduced the function
\beal
\label{eq:def_G}
\mathcal{G}(x)&=\frac{x\, e^x}{(e^x-1)^2}=x\,\nPl(x)[1+\nPl(x)] \equiv \frac{\tilde{S}^2}{4x}.
\end{align}
However, the above expressions by no means are unique. For example, one can also write
$\xg^3\partial^3_{\xg} \nPl =-\mathcal{G}\left[\frac{3}{2}\tilde{S}^2 +\xg^2 \right]$.
Since $\mathcal{G}$ is directly related to $\tilde{S}$ we therefore tried expressing everything with just $\tilde{X}$ and $\mathcal{G}$.
Using the relations
\bsub
\label{eq:deriv_G_A}
\beal
\xg\partial_{\xg}\mathcal{G}&=\mathcal{G}-\mathcal{G} \tilde{X}
\\
\xg\partial_{\xg}\tilde{X}&=\tilde{X}-2\xg\mathcal{G}
\\
\tilde{X}^2&=4\xg\mathcal{G}+\xg^2
\end{align}
\esub
it is straightforward to show that
\beal
\label{eq:relations_nPl}
\xg\partial_{\xg} \nPl &=-\mathcal{G},
\nonumber\\
\xg^2\partial^2_{\xg} \nPl &=+\mathcal{G} \tilde{X}
\nonumber\\
\xg^3\partial^3_{\xg} \nPl &=-\xg\mathcal{G}\left[6 \mathcal{G} +\xg \right]
\nonumber\\
\xg^4\partial^4_{\xg} \nPl &=+\xg\mathcal{G}\tilde{X}\left[12 \mathcal{G} +\xg \right]
\nonumber\\
\xg^5\partial^5_{\xg} \nPl &=-\xg\mathcal{G}
\left[30\mathcal{G}\tilde{X}^2 +\xg^3 \right]
\nonumber\\
\xg^6\partial^6_{\xg} \nPl &=+\xg^2\mathcal{G}\tilde{X}
\left[60\mathcal{G}\left(6\mathcal{G}+\xg\right) +\xg^2 \right]
\nonumber\\
\xg^7\partial^7_{\xg} \nPl &=-\xg^3\mathcal{G}
\left[42\mathcal{G}\left(40\mathcal{G}\left[3\mathcal{G}+\xg\right]+3\xg^2\right) +\xg^3 \right]
\nonumber\\
\xg^8\partial^8_{\xg} \nPl &=+\xg^2\mathcal{G}\tilde{X}
\left[252\mathcal{G}\left(20\mathcal{G}\tilde{X}^2+\xg^3\right) +\xg^4 \right].
\end{align}
Together with the moments, Table~\ref{tab:monopole}-\ref{tab:quadrupole}, these determine all temperature correction terms up to $\mathcal{O}(\The^{3})$.

\subsection{Closed form for the derivatives of a Planck spectrum}
\label{app:nPl_derives_closed}
In Eq.~\eqref{eq:relations_nPl} we give explicit expressions for the derivatives of the Planck spectrum up to the $8^{\rm th}$ derivative.
For fully analytic manipulations of the Boltzmann equation and when calculating the spectral distortion introduced by the scattering of photons off of electrons inside a moving cluster it is enough to have some way of computing the derivatives up to the desired order.
Looking at the main scaling at high frequencies one can show that $\partial^k_{\xg} \nPl \sim (-1)^k e^{-\xg} / (1-e^{-\xg})^{k+1}$.
In general, one can also write
\beal
\label{eq:deriv_nPl_euler}
\xg^k\partial^k_{\xg} \nPl &= \frac{(-\xg)^k e^{-\xg}}{(1-e^{-\xg})^{k+1}} \mathcal{P}_{k-1}(e^{-\xg}),
\end{align}
where $\mathcal{P}_k(x)$ is a polynomial of order $k$.

With the Ansatz $\mathcal{P}_k(x)=\sum^k_{m=0}\alpha^{(k)}_m x^m$ it is straightforward to find the recursion relation 
\beal
\label{eq:recursion_alpha}
\gamma^{(k)}_m &=(m+1)\, \gamma ^{(k-1)}_{m}+(k-m)\, \gamma ^{(k-1)}_{m-1}
\end{align}
for the coefficient $\gamma ^{(k)}_m$.
The solution of this recursion is
\beal
\label{eq:recursion_alpha_solution}
\gamma ^{(k)}_m &=\sum_{s=0}^{m} 
\,(-1)^s 
\left(\!\!
\begin{array}{c}
k+1 \\
s
\end{array}
\!\!\right) (m+1-s)^k
\equiv
\left<\!\!
\begin{array}{c}
k \\
m
\end{array}
\!\!\right>.
\end{align}
These coefficients are also known as Eulerian numbers. which can be generated in a very fast and numerically stable way using Eq.~\eqref{eq:recursion_alpha}.
The derivatives of the Planckian spectrum are given by
\beal
\label{eq:deriv_nPl_euler_final}
\xg^k\partial^k_{\xg} \nPl 
&= 
\frac{(-\xg)^k e^{-\xg}}{(1-e^{-\xg})^{k+1}} 
\sum_{m=0}^{k-1}\left<\!\!
\begin{array}{c}
k \\
m
\end{array}
\!\!\right> e^{-m\xg}
\end{align}
for any $k$. 
Defining $\mathcal{H}_k=(-1)^k\xg^k e^{-\xg}/(1-e^{-\xg})^{k+1}$ one can also write the derivatives as
\beal
\label{eq:relations_nPl_H}
\xg\partial_{\xg} \nPl &=\mathcal{H}_1,
\\
\xg^2\partial^2_{\xg} \nPl &=\mathcal{H}_2 \left[1+e^{-\xg}\right]
\nonumber\\
\xg^3\partial^3_{\xg} \nPl &=\mathcal{H}_3 \left[1+4e^{-\xg}+e^{-2\xg}\right]
\nonumber\\
\xg^4\partial^4_{\xg} \nPl &=\mathcal{H}_4 \left[1+11 e^{-\xg}+11 e^{-2\xg}+e^{-3\xg}\right]
\nonumber\\
\xg^5\partial^5_{\xg} \nPl &=\mathcal{H}_5 \left[1+26 e^{-\xg}+66 e^{-2\xg}+26e^{-3\xg}+e^{-4\xg}\right]
\nonumber\\
\xg^6\partial^6_{\xg} \nPl &=\mathcal{H}_6 
\left[1+57 e^{-\xg}+302 e^{-2\xg}
+302 e^{-3\xg}+57e^{-4\xg}+e^{-5\xg}\right].
\nonumber
\end{align}
As these expressions show, for $\xg\gg 1 $ the limiting behaviour is given by $\xg^k\partial^k_{\xg} \nPl \sim \mathcal{H}_k \sim (-1)^k\xg^k e^{-\xg}$.
Similarly, in the limit $\xg\ll 1 $ one has $\xg^k\partial^k_{\xg} \nPl \sim (-1)^k\xg^k$. 
For numerical applications these expressions are very useful and easy to code up.

To evaluate the sum $Y_m=\sum_{k=1}^{2(m+1)} a^{(m)}_k  \xg^k\partial^k_{\xg} \nPl$, which appears after collecting terms of different orders in $\The$ from the moments, $I^{k}_{lm}$, we can use the recursion relation
%
\beal
\label{eq:recursion_alpha_sums}
\beta^{(m)}_k &=\frac{-\xg}{1-e^{-\xg}} 
\left[ a^{(m)}_k \, \mathcal{P}_{k-1}(e^{-\xg}) +\beta^{(m)}_{k+1} \right]
\end{align}
with $\beta^{(m)}_{2(m+1)+1}=0$ and $Y_m(\xg)=\beta^{(m)}_1(\xg)\,\nPl(\xg)$. This groups terms of similar order and yields numerically stable results, even if the coefficients $a^{(m)}_k$ become very large (see Table~\ref{tab:monopole}).

\section{Azimuthally averaged cross sections}
\label{app:phi_av_sig}
The 5-dimensional collision integral can be reduced by 2 dimensions when considering the azimuthal symmetry of the scattering process. In particular, we encounter the integrals
\bsub
\beal
\label{eq:sig_0}
\frac{\id^2\sigma_0}{\id \mu \id\mu'}&\equiv 
\int\!\! \frac{\id\sigma}{\id \Omega'}\,\frac{\id\phi' \id\phi}{4\pi} 
\\
\nonumber
&=\frac{3\,\sigT}{8}\, \zeta^2 \gamma^2\kappa^3
\left[1-\zeta\left<\alpha_{\rm sc}\right>
+\frac{1}{2} \zeta^2
\left<\alpha_{\rm sc}^2\right>
\right]
\\[1mm]
\label{eq:sig_1}
\frac{\id^2\sigma_1}{\id \mu \id\mu'}&\equiv 
\int\!\! \mu_{\rm sc}\frac{\id\sigma}{\id \Omega'}\,\frac{\id\phi' \id\phi}{4\pi} 
\\
\nonumber
&=
\mu\mu'\frac{\id^2\sigma_0}{\id \mu \id\mu'}
+\frac{3\,\sigT}{8}\,\zeta^3 \gamma^2\kappa^3
\left[1
-\zeta\left<\alpha_{\rm sc}\right>
\right] g(\mu,\mu')
\\[1mm]
\label{eq:sig_2}
\frac{\id^2\sigma_2}{\id \mu \id\mu'}&\equiv 
\int\!\! P_2(\mu_{\rm sc})\frac{\id\sigma}{\id \Omega'}\,\frac{\id\phi' \id\phi}{4\pi} 
\\
\nonumber
&=
P_2(\mu\mu')\frac{\id^2\sigma_0}{\id \mu \id\mu'}
+\frac{15\,\sigT}{16}\,\zeta^4 \gamma^2\kappa^3\,[P_2(\mu\mu')-1]\,g(\mu,\mu')
\\
&\quad
+\frac{9\,\sigT}{16}\,\zeta^2 \gamma^2\kappa^3
\left[1+\zeta\left(2+\frac{3}{4}\zeta \,g-3\left<\alpha_{\rm sc}\right>[1-\zeta]\right)
\right] g
\nonumber
\end{align}
\esub
with $g(\mu,\mu')=\frac{1}{2}(1-\mu^2)(1-\mu'^2)$, $\zeta=\nu'/(\nu \gamma^2\kappa^2)$, and the averaged quantities $\left<\alpha_{\rm sc}\right>=\int (1-\mu_{\rm sc})\id\phi' \id\phi/4\pi^2=1-\mu\mu'$ and $\left<\alpha_{\rm sc}^2\right>=\int (1-\mu_{\rm sc})^2\id\phi' \id\phi/4\pi^2=\left<\alpha_{\rm sc}\right>^2+g(\mu,\mu')$.

\begin{table*}
\caption{Moments for monopole scattering. Blank means the coefficient is zero. Within each row, the temperature order increases from left ($\propto\The$) to right ($\propto\The^{11}$), while in each column the order of the derivative of the blackbody spectrum increases from top ($\xg \partial_{\xg} \nPl$) to bottom ($\xg^{22} \partial^{22}_{\xg} \nPl$). 
Each row therefore gives the coefficients $a_k^{(m)}$ for the moments $I^{k}_{00}=\The \sum_{m=0}^n a_k^{(m)} \The^{m} Y_{00}$, while each column defines the spectral function $Y_n=\sum_{k=1}^{2n+2} a_k^{(n)} \xg^k\partial^k_{\xg} n_{00}(\xg) Y_{00}$. We did not list the moment $I^{0}_{00}=Y_{00}=a_0^{(-1)} Y_{00}$ for Thomson scattering of the monopole in the table.}
\label{tab:monopole}
\centering
\begin{tabular}{@{}cccccccccccc}
& $Y_0$ & $Y_1$ & $Y_2$ & $Y_3$ & $Y_4$ & $Y_5$ 
& $Y_6$ & $Y_7$ & $Y_8$ & $Y_9$ & $Y_{10}$    
\\
\hline
\hline
$I^{1}_{00}$ & $4$ & $10$ & $\frac{15}{2}$ & $-\frac{15}{2}$ & \
$\frac{135}{32}$ & $\frac{45}{8}$ & $-\frac{7425}{256}$ & \
$\frac{675}{8}$ & $-\frac{1905525}{8192}$ & $\frac{91125}{128}$ & \
$-\frac{169255575}{65536}$
\\[0.8mm]
$I^{2}_{00}$ & $1$ & $\frac{47}{2}$ & $\frac{1023}{8}$ & \
$\frac{2505}{8}$ & $\frac{30375}{128}$ & $-\frac{7515}{32}$ & \
$\frac{128655}{1024}$ & $\frac{6345}{32}$ & $-\frac{31843125}{32768}$ \
& $\frac{1451925}{512}$ & $-\frac{2090024775}{262144}$
\\[0.8mm]
$I^{3}_{00}$ &  & $\frac{42}{5}$ & $\frac{868}{5}$ & $\frac{7098}{5}$ \
& $\frac{62391}{10}$ & $\frac{28917}{2}$ & $\frac{360675}{32}$ & \
$-\frac{86751}{8}$ & $\frac{1274427}{256}$ & $\frac{95823}{8}$ & \
$-\frac{433967625}{8192}$
\\[0.8mm]
$I^{4}_{00}$ &  & $\frac{7}{10}$ & $\frac{329}{5}$ & \
$\frac{14253}{10}$ & $\frac{614727}{40}$ & $\frac{795429}{8}$ & \
$\frac{50853555}{128}$ & $\frac{28579473}{32}$ & \
$\frac{723764619}{1024}$ & $-\frac{21387969}{32}$ & \
$\frac{8992650375}{32768}$
\\[0.8mm]
$I^{5}_{00}$ &  &  & $\frac{44}{5}$ & $\frac{18594}{35}$ & \
$\frac{124389}{10}$ & $\frac{2319993}{14}$ & $\frac{45719721}{32}$ & \
$\frac{463090581}{56}$ & $\frac{56131109271}{1792}$ & \
$\frac{3867907059}{56}$ & $\frac{3162934444995}{57344}$
\\[0.8mm]
$I^{6}_{00}$ &  &  & $\frac{11}{30}$ & $\frac{12059}{140}$ & \
$\frac{355703}{80}$ & $\frac{12667283}{112}$ & \
$\frac{458203107}{256}$ & $\frac{8680356807}{448}$ & \
$\frac{2131228533597}{14336}$ & $\frac{361018793313}{448}$ & \
$\frac{1353148643034945}{458752}$
\\[0.8mm]
$I^{7}_{00}$ &  &  &  & $\frac{128}{21}$ & $\frac{16568}{21}$ & \
$\frac{806524}{21}$ & $\frac{22251961}{21}$ & $\frac{407333911}{21}$ \
& $\frac{28385005515}{112}$ & $\frac{68346357865}{28}$ & \
$\frac{15614127041155}{896}$
\\[0.8mm]
$I^{8}_{00}$ &  &  &  & $\frac{16}{105}$ & $\frac{7516}{105}$ & \
$\frac{21310}{3}$ & $\frac{71548297}{210}$ & $\frac{304758409}{30}$ & \
$\frac{33759855933}{160}$ & $\frac{129419653687}{40}$ & \
$\frac{9649428040913}{256}$
\\[0.8mm]
$I^{9}_{00}$ &  &  &  &  & $\frac{22}{7}$ & $\frac{46679}{63}$ & \
$\frac{26865067}{420}$ & $\frac{14281971623}{4620}$ & \
$\frac{7332664403233}{73920}$ & $\frac{14191595238489}{6160}$ & \
$\frac{1600697595147911}{39424}$
\\[0.8mm]
$I^{10}_{00}$ &  &  &  &  & $\frac{11}{210}$ & $\frac{10853}{252}$ & $\
\frac{7313155}{1008}$ & $\frac{32154229291}{55440}$ & \
$\frac{8454102129551}{295680}$ & $\frac{72688716977749}{73920}$ & \
$\frac{11942478518370683}{473088}$
\\[0.8mm]
$I^{11}_{00}$ &  &  &  &  &  & $\frac{58}{45}$ & $\frac{4492}{9}$ & \
$\frac{34276642}{495}$ & $\frac{159273899}{30}$ & \
$\frac{1153370108027}{4290}$ & $\frac{135515512037513}{13728}$
\\[0.8mm]
$I^{12}_{00}$ &  &  &  &  &  & $\frac{29}{1890}$ & $\frac{6361}{315}$ \
& $\frac{36841447}{6930}$ & $\frac{3623853049}{5544}$ & \
$\frac{393223901251}{8008}$ & $\frac{985744536107759}{384384}$
\\[0.8mm]
$I^{13}_{00}$ &  &  &  &  &  &  & $\frac{296}{675}$ & \
$\frac{1927084}{7425}$ & $\frac{133798003}{2475}$ & \
$\frac{197855054569}{32175}$ & $\frac{330716455601941}{720720}$
\\[0.8mm]
$I^{14}_{00}$ &  &  &  &  &  &  & $\frac{37}{9450}$ & \
$\frac{229693}{29700}$ & $\frac{118712629}{39600}$ & \
$\frac{275619041167}{514800}$ & $\frac{667649235923203}{11531520}$
\\[0.8mm]
$I^{15}_{00}$ &  &  &  &  &  &  &  & $\frac{736}{5775}$ & \
$\frac{631168}{5775}$ & $\frac{813344128}{25025}$ & \
$\frac{3734552464}{715}$
\\[0.8mm]
$I^{16}_{00}$ &  &  &  &  &  &  &  & $\frac{46}{51975}$ & \
$\frac{1312}{525}$ & $\frac{102350176}{75075}$ & \
$\frac{1173672524}{3465}$
\\[0.8mm]
$I^{17}_{00}$ &  &  &  &  &  &  &  &  & $\frac{16}{495}$ & \
$\frac{248008}{6435}$ & $\frac{101283718}{6435}$
\\[0.8mm]
$I^{18}_{00}$ &  &  &  &  &  &  &  &  & $\frac{4}{22275}$ & \
$\frac{22448}{32175}$ & $\frac{46573313}{90090}$
\\[0.8mm]
$I^{19}_{00}$ &  &  &  &  &  &  &  &  &  & $\frac{268}{36855}$ & \
$\frac{231496}{19845}$
\\[0.8mm]
$I^{20}_{00}$ &  &  &  &  &  &  &  &  &  & $\frac{67}{2027025}$ & \
$\frac{486254}{2837835}$
\\[0.8mm]
$I^{21}_{00}$ &  &  &  &  &  &  &  &  &  &  & $\frac{632}{429975}$
\\[0.8mm]
$I^{22}_{00}$ &  &  &  &  &  &  &  &  &  &  & $\frac{79}{14189175}$
\\
\hline
\hline
\end{tabular}
\end{table*}

\begin{table*}
\caption{Moments for dipole scattering. The table is organized similar to Table~\ref{tab:monopole}.
Each row gives the coefficients $d_k^{(m)}$ for the moments $I^{k}_{1m}= \The \sum_{j=0}^n d_k^{(j)} \The^{j} Y_{1m}$, while each column defines the spectral function $D_n=\sum_{k=0}^{2n+2} d_k^{(n)} \xg^k\partial^k_{\xg} n_1(\xg)$, with $n_1(\xg)\equiv \sum_{m} n_{1m}(\xg) Y_{1m}$. Since the dipole does not couple directly to the Thomson scattering cross section, all moments are at least first order in temperature.}
\label{tab:dipole}
\centering
\begin{tabular}{@{}cccccccccccc}
& $D_0$ & $D_1$ & $D_2$ & $D_3$ & $D_4$ & $D_5$ 
& $D_6$ & $D_7$ & $D_8$     
\\
\hline
\hline
$I^{0}_{1m}$ & $-\frac{2}{5}$ & $-\frac{1}{5}$ & $\frac{407}{140}$ & \
$-\frac{363}{28}$ & $\frac{23363}{448}$ & $-\frac{24895}{112}$ & \
$\frac{41394345}{39424}$ & $-\frac{6794331}{1232}$ & \
$\frac{523431480345}{16400384}$
\\[0.8mm]
$I^{1}_{1m}$ & $-\frac{8}{5}$ & $-\frac{24}{5}$ & $-\frac{233}{35}$ & \
$\frac{117}{7}$ & $-\frac{6077}{112}$ & $\frac{6145}{28}$ & \
$-\frac{10205655}{9856}$ & $\frac{1685589}{308}$ & \
$-\frac{130381012455}{4100096}$
\\[0.8mm]
$I^{2}_{1m}$ & $-\frac{2}{5}$ & $-\frac{66}{5}$ & \
$-\frac{10433}{140}$ & $-\frac{5463}{28}$ & $-\frac{51917}{448}$ & \
$\frac{3397}{112}$ & $\frac{17610897}{39424}$ & $-\frac{883335}{308}$ \
& $\frac{270896744505}{16400384}$
\\[0.8mm]
$I^{3}_{1m}$ &  & $-\frac{24}{5}$ & $-\frac{3876}{35}$ & \
$-\frac{32554}{35}$ & $-\frac{873109}{210}$ & $-\frac{403727}{42}$ & \
$-\frac{18930097}{2464}$ & $\frac{5023397}{616}$ & \
$-\frac{2222982373}{256256}$
\\[0.8mm]
$I^{4}_{1m}$ &  & $-\frac{2}{5}$ & $-\frac{1513}{35}$ & \
$-\frac{68869}{70}$ & $-\frac{9083573}{840}$ & \
$-\frac{11869231}{168}$ & $-\frac{2793622481}{9856}$ & \
$-\frac{1573137659}{2464}$ & $-\frac{515989449221}{1025024}$
\\[0.8mm]
$I^{5}_{1m}$ &  &  & $-\frac{204}{35}$ & $-\frac{13166}{35}$ & \
$-\frac{1903991}{210}$ & $-\frac{25726409}{210}$ & \
$-\frac{13112497879}{12320}$ & $-\frac{19051896541}{3080}$ & \
$-\frac{6016818388583}{256256}$
\\[0.8mm]
$I^{6}_{1m}$ &  &  & $-\frac{17}{70}$ & $-\frac{25903}{420}$ & \
$-\frac{16667831}{5040}$ & $-\frac{432370409}{5040}$ & \
$-\frac{405834499159}{295680}$ & $-\frac{1105432084861}{73920}$ & \
$-\frac{708397495533863}{6150144}$
\\[0.8mm]
$I^{7}_{1m}$ &  &  &  & $-\frac{92}{21}$ & $-\frac{37376}{63}$ & \
$-\frac{9325552}{315}$ & $-\frac{136723936}{165}$ & \
$-\frac{17674562992}{1155}$ & $-\frac{603792624896}{3003}$
\\[0.8mm]
$I^{8}_{1m}$ &  &  &  & $-\frac{23}{210}$ & $-\frac{17047}{315}$ & $-\
\frac{3483659}{630}$ & $-\frac{2494267697}{9240}$ & \
$-\frac{75202564463}{9240}$ & $-\frac{327273403197989}{1921920}$
\\[0.8mm]
$I^{9}_{1m}$ &  &  &  &  & $-\frac{50}{21}$ & $-\frac{20303}{35}$ & \
$-\frac{708479411}{13860}$ & $-\frac{11566377071}{4620}$ & \
$-\frac{77893231780421}{960960}$
\\[0.8mm]
$I^{10}_{1m}$ &  &  &  &  & $-\frac{5}{126}$ & $-\frac{213061}{6300}$ \
& $-\frac{179560553}{30800}$ & $-\frac{131179967167}{277200}$ & \
$-\frac{90620456461739}{3843840}$
\\[0.8mm]
$I^{11}_{1m}$ &  &  &  &  &  & $-\frac{76}{75}$ & \
$-\frac{331988}{825}$ & $-\frac{140555494}{2475}$ & \
$-\frac{6300995377}{1430}$
\\[0.8mm]
$I^{12}_{1m}$ &  &  &  &  &  & $-\frac{19}{1575}$ & \
$-\frac{847771}{51975}$ & $-\frac{151586069}{34650}$ & \
$-\frac{196409497183}{360360}$
\\[0.8mm]
$I^{13}_{1m}$ &  &  &  &  &  &  & $-\frac{2632}{7425}$ & \
$-\frac{1589204}{7425}$ & $-\frac{1455116041}{32175}$
\\[0.8mm]
$I^{14}_{1m}$ &  &  &  &  &  &  & $-\frac{47}{14850}$ & \
$-\frac{1327541}{207900}$ & $-\frac{9057624841}{3603600}$
\\[0.8mm]
$I^{15}_{1m}$ &  &  &  &  &  &  &  & $-\frac{608}{5775}$ & \
$-\frac{2296512}{25025}$
\\[0.8mm]
$I^{16}_{1m}$ &  &  &  &  &  &  &  & $-\frac{38}{51975}$ & \
$-\frac{12128}{5775}$
\\[0.8mm]
$I^{17}_{1m}$ &  &  &  &  &  &  &  &  & $-\frac{136}{5005}$
\\[0.8mm]
$I^{18}_{1m}$ &  &  &  &  &  &  &  &  & $-\frac{34}{225225}$
\\
\hline
\hline
\end{tabular}
\end{table*}

\begin{table*}
\caption{Moments for quadrupole scattering. The table is organized similar to Table~\ref{tab:monopole}, but we also add the coefficient $q_0^{(-1)}=1/10$ to the table, which arises from Thomson scattering of the quadrupole anisotropy and is independent of temperature.
Each row gives the moment coefficients $q_k^{(m)}$ for $I^{k}_{2m}= \The \sum_{j=-1}^n q_k^{(j)} \The^{j} Y_{2m}$, while each column defines the spectral function $Q_n=\sum_{k=0}^{2n+2} q_k^{(n)} \xg^k\partial^k_{\xg} n_2(\xg)$, with $n_2(\xg)\equiv \sum_{m=-2}^2 n_{2m}(\xg) Y_{2m}$.}
\label{tab:quadrupole}
\centering
\begin{tabular}{@{}ccccccccccccc}
& $Q_{-1}$ & $Q_0$ & $Q_1$ & $Q_2$ & $Q_3$ & $Q_4$ & $Q_5$ 
& $Q_6$ & $Q_7$ & $Q_8$     
\\
\hline
\hline
$I^{0}_{2m}$ & $\frac{1}{10}$ & $-\frac{3}{5}$ & $\frac{183}{70}$ & $-\frac{429}{40}$ \
& $\frac{2559}{56}$ & $-\frac{26961}{128}$ & $\frac{2619681}{2464}$ & \
$-\frac{462951045}{78848}$ & $\frac{1130474547}{32032}$ & \
$-\frac{7529179952565}{32800768}$
\\[0.8mm]
$I^{1}_{2m}$ && $\frac{2}{5}$ & $-\frac{5}{7}$ & $\frac{207}{20}$ & $-\
\frac{1269}{28}$ & $\frac{13467}{64}$ & $-\frac{1310187}{1232}$ & \
$\frac{231532695}{39424}$ & $-\frac{565304841}{16016}$ & \
$\frac{3764780719335}{16400384}$
\\[0.8mm]
$I^{2}_{2m}$ && $\frac{1}{10}$ & $\frac{115}{28}$ & $\frac{1647}{80}$ \
& $\frac{9531}{112}$ & $-\frac{14757}{256}$ & $\frac{2389605}{4928}$ \
& $-\frac{459217161}{157696}$ & $\frac{1133285355}{64064}$ & \
$-\frac{7542692912025}{65601536}$
\\[0.8mm]
$I^{3}_{2m}$ &&  & $\frac{12}{7}$ & $44$ & $\frac{13306}{35}$ & \
$\frac{18041}{10}$ & $\frac{610943}{154}$ & $\frac{10329129}{2464}$ & \
$-\frac{72011865}{8008}$ & $\frac{10182974397}{256256}$
\\[0.8mm]
$I^{4}_{2m}$ &&  & $\frac{1}{7}$ & $19$ & $\frac{32341}{70}$ & \
$\frac{209617}{40}$ & $\frac{21549415}{616}$ & \
$\frac{1391734017}{9856}$ & $\frac{10278017103}{32032}$ & \
$\frac{248720650869}{1025024}$
\\[0.8mm]
$I^{5}_{2m}$ &&  &  & $\frac{92}{35}$ & $\frac{6614}{35}$ & \
$\frac{66821}{14}$ & $\frac{7275659}{110}$ & \
$\frac{7186785003}{12320}$ & $\frac{19550802003}{5720}$ & \
$\frac{477065269101}{36608}$
\\[0.8mm]
$I^{6}_{2m}$ &&  &  & $\frac{23}{210}$ & $\frac{13387}{420}$ & \
$\frac{3067801}{1680}$ & $\frac{129481163}{2640}$ & \
$\frac{78933400569}{98560}$ & $\frac{404328355121}{45760}$ & \
$\frac{20084725151311}{292864}$
\\[0.8mm]
$I^{7}_{2m}$ &&  &  &  & $\frac{16}{7}$ & $336$ & \
$\frac{20263976}{1155}$ & $\frac{16645526}{33}$ & \
$\frac{142242005782}{15015}$ & $\frac{1007702100971}{8008}$
\\[0.8mm]
$I^{8}_{2m}$ &&  &  &  & $\frac{2}{35}$ & $\frac{1086}{35}$ & \
$\frac{3867887}{1155}$ & $\frac{780884497}{4620}$ & \
$\frac{312667550803}{60060}$ & $\frac{35387820016631}{320320}$
\\[0.8mm]
$I^{9}_{2m}$ &&  &  &  &  & $\frac{48}{35}$ & $\frac{1232564}{3465}$ & \
$\frac{37655546}{1155}$ & $\frac{2236313242}{1365}$ & \
$\frac{6483638594513}{120120}$
\\[0.8mm]
$I^{10}_{2m}$ &&  &  &  &  & $\frac{4}{175}$ & $\frac{361639}{17325}$ \
& $\frac{130359193}{34650}$ & $\frac{12883109639}{40950}$ & \
$\frac{7686639348727}{480480}$
\\[0.8mm]
$I^{11}_{2m}$ &&  &  &  &  &  & $\frac{1552}{2475}$ & \
$\frac{25892}{99}$ & $\frac{1227837074}{32175}$ & \
$\frac{18203744063}{6006}$
\\[0.8mm]
$I^{12}_{2m}$ &&  &  &  &  &  & $\frac{388}{51975}$ & \
$\frac{7375}{693}$ & $\frac{1333611599}{450450}$ & \
$\frac{136419177953}{360360}$
\\[0.8mm]
$I^{13}_{2m}$ &&  &  &  &  &  &  & $\frac{344}{1485}$ & \
$\frac{2808716}{19305}$ & $\frac{7122939553}{225225}$
\\[0.8mm]
$I^{14}_{2m}$ &&  &  &  &  &  &  & $\frac{43}{20790}$ & \
$\frac{2352131}{540540}$ & $\frac{44543067553}{25225200}$
\\[0.8mm]
$I^{15}_{2m}$ &&  &  &  &  &  &  &  & $\frac{5392}{75075}$ & \
$\frac{33981568}{525525}$
\\[0.8mm]
$I^{16}_{2m}$ &&  &  &  &  &  &  &  & $\frac{337}{675675}$ & \
$\frac{778982}{525525}$
\\[0.8mm]
$I^{17}_{2m}$ &&  &  &  &  &  &  &  &  & $\frac{6052}{315315}$
\\[0.8mm]
$I^{18}_{2m}$ &&  &  &  &  &  &  &  &  & $\frac{1513}{14189175}$\\
\hline
\hline
\end{tabular}
\end{table*}

\section{Moment for monopole, dipole and quadrupole scattering} 
\label{app:moments}
We are interested in kinematic corrections up to second order of the peculiar velocity. To this order, the CMB spectrum has a motion-induced dipole and quadrupole component inside the moving frame.
In $n^{\rm th}$ order of $\The$ we therefore need all moments $I^k_{00}$, $I^k_{1m}$, and $I^k_{2m}$ for $s\leq 2n+2$.
The required integrations were performed with the symbolic algebra package {\sc Mathematica}, defining appropriate replacement rules for all the appearing integrals in the Taylor expansion for small temperature.
Note that for the Taylor expansion of the relativistic Maxwell-Boltzmann distribution function, Eq.~\eqref{eq:relMBD}, it is useful to transform to the variable $\xi =\eta^2/2\The$.

The moments related to the monopole part of the photon distribution are summarized in Table~\ref{tab:monopole} for up to $10^{\rm th}$ order temperature corrections, or in total $11^{\rm th}$ order in temperature.
Up to $5^{\rm th}$ order in temperature these expressions agree with the results given by \citet{Itoh98} when neglecting all recoil terms.

The Tables~\ref{tab:dipole} and \ref{tab:quadrupole} give the moments for the scattering of the dipolar and quadrupolar spectral anisotropy in the radiation field, but only up to $\mathcal{O}(\The^9)$, for illustration. 
In {\sc SZpack} we included tables up to $10^{\rm th}$ order, similar to the monopole scattering case. 
The moments up to the first order in temperature were also derived in \citet{Chluba2012}.
However, higher moments have not been given in the literature before. 
In this work, we use then to write down all kinematic correction terms up to second order of the peculiar velocity using explicit Lorentz transformations.

\end{appendix}

\footnotesize
\bibliographystyle{mn2e}
\bibliography{Lit}
\end{document}